\documentclass[10pt]{article}

\usepackage[a4paper, left=3.5cm, right=3.5cm]{geometry}

\usepackage[utf8]{inputenc}
\usepackage[T1]{fontenc}
\usepackage{graphicx}
\usepackage{longtable}
\usepackage{wrapfig}
\usepackage{caption}
\usepackage{subcaption}

\usepackage{url}
\usepackage{hyperref} 

\usepackage{amsmath,amssymb,bm}
\usepackage{authblk}

\usepackage{hyperref}
\usepackage{svg}
\usepackage{float}
\usepackage{multirow}

\usepackage{setspace}

\date{\today}

\hypersetup{
 pdfauthor={B. Barzdajn},
 pdftitle={Statistical modelling of irradiation damage and potential applications},
 pdfkeywords={irridiation damage, Zr, primary knock-on atom, collision cascade},
 pdfsubject={scientific paper},
 pdfcreator={pdflatex}, 
 pdflang={English}}

\begin{document}

\newcommand{\cmt}[1]{}
\newcommand{\note}[1]{{\color{red} #1}}
\newcommand{\ts}[1]{\textsuperscript{#1}}
\newcommand{\txt}[1]{\mathrm{#1}}

\title{Molecular dynamics simulations of neutron induced collision cascades in Zr -- statistical modelling of irradiation damage and potential applications}

\author{Bartosz Barzdajn\footnote{bartosz.barzdajn@manchester.ac.uk}}
\affil{The University of Manchester, Department of Materials}
\author{Christopher P Race}
\affil{The University of Sheffield, Department of Materials Science and Engineering}

\date{\today}
\maketitle

\begin{abstract} 
Understanding the nature of irradiation damage often requires a multi-scale and multi-physics approach, \textit{i.e.} it requires a significant amount of information from experiments, simulations and phenomenological models. This paper focuses on the initial stages of irradiation damage, namely neutron-induced displacement cascades in zirconium, as nuclear-grade zirconium alloys are widely used in fuel assemblies. We provide results of large-scale molecular dynamics (MD) simulations based on existing inter-atomic potentials and the two-temperature model to include the effect of electron-phonon coupling. Our data can be used directly in higher scale methods. Furthermore, we analysed summary statistics associated with defect production, such as the number of defects produced, their distribution and the size of clusters. As a result, we have developed a generative model of collision cascades. The model is hierarchical, as well as stochastic, \textit{i.e.} it includes the variance of the considered features. This development had three main objectives: to establish a sufficient descriptor of a cascade, to develop an interpolator of data obtained from high-fidelity simulations, and to demonstrate that the statistical model of the data can generate representative distributions of primary irradiation defects. The results can be used to generate synthetic inputs for longer length- and time-scale models, as well as to build fast approximations relating dose, damage and irradiation conditions.
\end{abstract}

\section{Introduction}
\label{sec:org48e4150}

One of the biggest challenges in the safe and efficient design and operation of nuclear reactors, is achieving a deep understanding of changes in materials properties under the extreme conditions of the reactor core.

The subject of our investigation is zirconium. Nuclear grade Zr alloys are widely used in fuel assemblies due to their low thermal neutron absorption and good corrosion resistance. However, they exhibit irradiation-induced growth (anisotropic and near-volume conserving changes of shape) and creep \cite{adamson_irradiation_2019}. This can pose significant problems for structural engineers even under the assumption of normal operational conditions.

Our research focuses on the initial stages of neutron radiation damage, namely neutron-induced displacement cascades in zirconium. We use existing empirical potentials to perform large scale simulations and quantify the distribution of resulting defect populations.

The phenomenon in question can be briefly summarised as follows. A portion of the energy of the incident neutron is transferred in the form of kinetic energy to one of the atoms constituting the solid. This is the primary knock-on atom (PKA). This atom initiates a series of collisions cascade, generating a localised region of high energy atoms (with effective temperature of a several thousand Kelvin) leaving behind a cluster of point defects consisting of dozens of pairs of vacancies and self-interstitial atoms (Frenkel pairs). While a single cascade last for tens of  \(\mathrm{ps}\), there are billions or trillions of cascades per \(\mathrm{cm^3}\), depending on the neutron energy and type of the reactor, happening each second. Some of the resulting defects will be annealed-out, but others might give rise to the formation of clusters, dislocation loops or voids.

The dynamics of these defects will be responsible for changes in mechanical properties after prolonged exposures. This creates a truly multiscale problem that involves scales from $\txt{nm}$ to $\txt{m}$ and time scales from $\txt{ps}$ to years.

Since the primary stages of radiation damage are inaccessible for experimentalists, there are many examples of simulation-based studies in the literature. Something that could be referred to as a modern approach, that involves explicit simulation of the phenomenon at the atomistic scale can be dated back to the classical works of Bacon \textit{et al.} \cite{bacon_molecular_1994,bacon_computer_1995,bacon_defect_1997}. The primary focus was then on a deeper understanding and realistic quantification of the defect production process that goes beyond the simple NRT model \cite{norgett_proposed_1975} or the binary collision approximation \cite{bohr_bca_1948,robinson_binary_1994}.

In the context of Zr, further developments include, among others, consideration of a wider range of simulation parameters and methods to summarise outcomes, including various temperatures and PKA energies \cite{voskoboinikov_statistics_2006}, strain effects \cite{di_molecular_2013,sahi_molecular_2018}, simulations of high energy PKAs \cite{zhou_molecular_2018}, inclusion of electron-phonon coupling \cite{zhou_effects_2020}, etc.


While there are many others examples of such studies in the literature, there are reasons to undertake further numerical experiments. Firstly, ever-increasing computational power allowed us to obtain large, high-quality data sets, designed with a carefully selected sample of the primary knock-on atom initial momenta. At the same time, we were able to extend the range of phenomena accounted for in the simulations to include the electronic stopping and electron-ion interactions via the two-temperature MD model (TTM, \cite{duffy_including_2007,rutherford_effect_2007}).


We adopt the argument made by Bacon and Rubia \cite{bacon_molecular_1994}, that a deep understanding requires not only recognition of phenomena, but also quantitative assessment. This is particularly important for smaller space and time scales that inform other methods. We argue that this must go beyond simple point estimates. Most researchers focus on reporting best estimates e.g. for the number of point defects produced in a cascade. However, we argue that for a predictive framework (quantitive predictions) the description of a phenomenon must include information about associated probability distributions.

Hence, we focus here on the development of a generative model for collision cascades. We will present a framework that takes the momentum of a primary knock-on atom as input and produces a statistically representative set of characteristics of the resulting cascade damage, namely the number and spatial distribution of defects produced.

Such a model can be used directly to generate defects for higher scale methods, such as kinetic/object Monte Carlo -- kMC/oMC (\cite{nordlund_historical_2019}) or to support other methods which rely on introduction of defects to the system such as, the creation-relaxation algorithm -- CRA \cite{derlet_microscopic_2020,warwick_microstructural_2021}, or random displacement approximation -- RDA \cite{maxwell_atomistic_2020} \cmt{This is a long shot ... I'm not sure how they work.}. In other words, our model can be used to interpolate databases of collision cascades. Our approach also provides insights into which characteristics of cascades affect the damage produced and how those characteristics should be reported.

Finally, we will address how predicted damage can be linked with results from other studies with scales that could reach to measurable quantities. \cmt{For example, ... We need to discuss this.}

\subsection{Benefits a generative model for collision cascades}
\label{sec:orga96c867}

In this section, we will elaborate on the arguments presented in the introduction. Each collision cascade can be considered as essentially unique, given the huge range of potential outcomes. At elevated temperatures, nominally equivalent PKA’s, with the same momentum, can produce significantly different collections of Frenkel pairs.

To pass information about defect production to a higher time and length scale method we could simply use direct simulation to generate a large sample of possible outcomes. However, the predictive value of a higher scale method (\textit{e.g.} kMC, RDA/CRA) will improve if it takes into account realistic PKA energy spectra representing reactor core conditions. Valid representation of a full PKA spectrum might require a large and expensively acquired simulation database Throughout, we make the assumption that local spatial correlations in positions of defects, rather than just defect numbers, are important.

The key factors in comparing a generative model with a direct sampling from a simulation database are as follows.

\begin{enumerate}
\item Appropriate direct sampling of the energy and momentum spectrum for the PKA will require a database consisting of hundreds or even thousands of simulations. We need to also consider that such a database will require a dense sampling of a wide range of energies to be applicable in a wide range of cases. With more accurate implementations of interatomic potentials, such as Gaussian approximation potential (GAP, \cite{bartok_gaussian_2010}) or other machine-learning (ML) potentials, the construction of a comprehensive database can be even more computationally demanding.

\item\label{pow_to_explain} Another advantage of a generative model is that it can highlight which characteristics of collision cascades, e.g., size, density etc., suffice for an appropriate description. In other words, we can compare defects generated using the model with the results of explicit MD simulations to verify and optimise features selected to be inputs.

\item A database of MD simulations will have a limited coverage and will be generated for specific conditions like temperature or PKA spectra. Some situations require an increase of the sampling density within a certain range of conditions (interpolation) or an expansion of this range (extrapolation). When the software used to generate the data is unavailable, or we are missing some essential information about the data-generating process, it might be impossible. On the other hand, the parameters of generative models can be extrapolated and interpolated. For example, we can use discrete data of defect production efficiency as a function of hydrostatic strain (\textit{e.g.} \cite{sahi_molecular_2018}) and adjust the generating algorithm to provide a continuous approximation.

\item\label{feed_forward} Generative models can be used to build simple simulations which could serve as baseline/initial approximations of important phenomena that would be extremely expensive to simulate explicitly. For example, in a simplified experiment where defects are created according to estimated sampling distributions, and existing defects are annealed when they overlap with later cascades, we can relate the number of defects introduced into a material to the number of incidents. This can be used \textit{e.g.} to build a simple approximation to a mapping between dose (energy transferred) and canonical dpa (\textit{i.e.} number of defects produced, used \textit{e.g.} in \cite{warwick_microstructural_2021}). However, such a mapping will depend on the PKA spectrum. We will present such an experiment 

\item More generally, with a statistical model that can generate representative defect populations we can use a network of conditional probabilities to inform experiments and higher-scale methods. For example, we could infer the dose from the defect density and PKA spectrum. When we think about the flow of information, this is an inverse of the case \ref{feed_forward} considered above. However, this specific example points towards an even more general challenge, namely the implementation of Bayesian modelling in the context of multi-scale and multi-physics frameworks.

\item Treating the results of simulations and experiments as distributions allows us to link information gathered from many diverse sources of information. Even more generally, we think that summarising results using generative models brings us one step closer to combining multiple studies into a quantitive predictive framework.
\end{enumerate}

The following sections will consider the methodology for building a database that will underlie the statistical model, followed by our approach to building a fit-for-purpose descriptor and associated predictive distributions. Together they will form the generative model of defect populations produced in a collision cascade.

\section{Methods and techniques}
\label{sec:methods}

To summarise our methodology in a few sentences, we use molecular dynamics (MD) to simulate explicitly the process of forming defects as a result of collision cascades. We repeat this for a variety of initial conditions. The resulting populations of point defects will be summarised as statistical distributions, which will be used to generate representative results without the need for further simulations. These distributions will be conditioned on the characteristics of the primary-knock on atoms (PKA's) and environmental conditions.

We use LAMMPS (stable release 3 March 2020, \cite{LAMMPS}) for MD simulations and the following Python packages for pre- and post-processing: ASE \cite{larsen_atomic_2017}, Atomman \cite{atomman}, NumPy \cite{harris2020array}, Pandas \cite{reback2020pandas,mckinney-proc-scipy-2010}, Matplotlib \cite{Hunter:2007}, Seaborn \cite{Waskom2021} and Ovito \cite{ovito}.

There are two main aspects of the creation of data for the generative model: a sampling of the PKA phase space and the underlying physical model of the simulations. First, we address the sampling.

In ``classical'' simulations, where quantum-mechanical interactions are approximated by an interatomic potential, the complete information about the system, a perfect hcp Zr lattice in our case, consists of positions and momenta of all atoms. We aim to maximise the efficiency and minimise the number of simulations used to estimate the properties of the distributions. Hence, we introduce a sampling strategy inspired by importance sampling for a uniform distribution.

Simulations of cascades consisted of two stages. The first one we call the geometry-optimisation stage in which we introduced atomic velocities (with average kinetic energy that corresponds to $1200 \, \txt{K}$, Gaussian distribution), thermalise the system using the NVT ensemble to $600 \, \txt{K}$ (Nose-Hover style equations of motion, \cite{lammps_fix_NH,shinoda_rapid_2004}), optimised the geometry using the NPH ensemble to zero pressure and $600 \, \txt{K}$ (same equations of motion as before only with the Langevin thermostat, \cite{lammps_fix_lang,schneider_molecular-dynamics_1978}) and normalised the distribution using NVE ensemble. The resulting state we regard as a good and universal starting point for the following simulations.

Prior to introducing excess PKA momentum, the system was again thermalised to a temperature of 600 K using the NVT assemble. This temperature is representative of the realistic operating conditions of a pressurised water reactor (PWR) core. Since we optimised the shape of the computational cell in the previous stage, the pressure oscillated near zero. Each thermalisation was conducted using different random seeds of the thermostat. This way, each simulation starts with a unique configuration of all atoms, but in a equivalent thermodynamic state. 

Next, the centre atom, designated as the primary knock-on atom (PKA), was given an initial momentum from a predefined set. The set was prepared in two stages. First, we randomly and uniformly generate the required number points on the surface of a unit sphere \cite{wolf_sphere_point_pick}. Next, we solve the Thomson problem, \textit{i.e.} we optimise positions on the surface by minimising an arbitrary inverse-distance potential. From the position we extract directions of the momenta. The magnitude is obtained from uniform sampling of the kinetic energy, hence the sampling will be more concentrated for higher momenta. 

To set the sampling range, we analysed PWR PKA spectra (we used the spectrum for Zr as reported in \cite{noori-kalkhoran_evaluation_2020}) and decided to focus on the lower half, i.e. the maximum energy considered corresponds to the median of the distribution -- approx $40\, \txt{keV}$ . The reasoning is that higher energy cascades tend to break into sub-cascades. This is a common assumption justifying the focus on lower energy cascades. However, as reported by Zhou \textit{et al.} in \cite{zhou_molecular_2018}, in reality high-energy cascades will branch-out, although a significant portion of them will form connected cascades. On the other hand, Zhou \textit{et al.} also found that for PKAs above $40\, \txt{keV}$ only a small portion of cascades could be classified as unfragmented.

\begin{figure}[h!]
\centering
\includegraphics[width=1.0\textwidth]{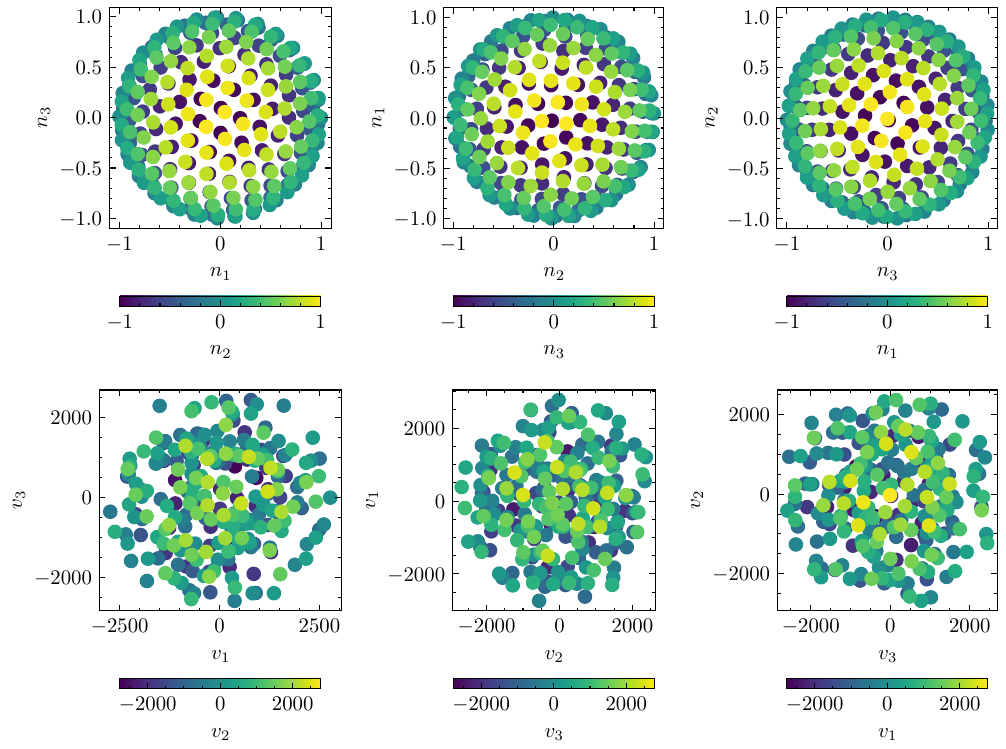} 
\caption{The sample of 256 initial directions of PKAs ($\vec{n}$, top row) and their velocities ($\vec{v}$, bottom row) given in \(\mathrm{\mathring{A} / ps}\).}
\label{fig:pka_momenta}
\end{figure}

To summarise, our simulations, involve approximately 2 million Zr atoms at 600 K. We use optimised uniform sampling of PKA direction and uniform sampling of its kinetic energy. Such sampling allows us to easily propagate results (\textit{e.g} the number of defects created per cascade) for any distribution/spectrum using an inverse probability integral transform. The resulting set of PKA directions and velocities is presented in Figure \ref{fig:pka_momenta}.

Now we consider the simulation method in greater detail. As mentioned before, we employ classical molecular dynamics. This means complex quantum-mechanical interactions between electrons and nuclei are abstracted into a numerical or closed-form potential that depends on the description of the local chemical environment (relative coordinates of surrounding atoms). We use the embedded atom method (EAM, \cite{daw_embedded-atom_1993}), where the chemical environment for each atom consists of a set of pair-wise distances to atoms within a specified radius and a local electronic density to which nearby atoms contribute.

We adopt Mendelev and Ackland's parametrisation often designated as the potential MA\#2 \cite{mendelev_development_2007}. The most commonly used is potential MA\#3, mainly due to better-fitted vacancy and interstitial formation energies. However, unlike the MA\#3, the selected MA\#2 potential predicts a non-decreasing relationship between temperature and lattice parameter (under constant c/a ratio), which is a desirable characteristic in high temperature simulations.

All MA potentials are consistent with the Ziegler-Biersack-Littmark (ZBL) potential \cite{zbl_pair_style_lammps,ziegler1985stopping}. The latter is considered a very good description of close collisions that will dominate the ballistic phase of the system evolution.

In our studies we placed a significant emphasis on the management of the heat transfer. Since molecular dynamics simulations do not explicitly consider electrons we employ the two-temperature model (TTM) developed by D. M. Duffy and A. M. Rutherford \cite{duffy_including_2007,rutherford_effect_2007}. This model incorporates redistribution of heat by electrons and transfer between electronic and atomic subsystems. This phenomenon can be also referred to as electron-phonon coupling. It is an essential part of the model as defects are formed during recovery after a heat spike that raises local temperature above the melting point. As demonstrated in \cite{rutherford_effect_2007}, coupling effects can reduce the number of defects created due to enhanced energy/heat transfer through the electronic subsystem. Electronic heat transfer dominates phonon propagation which occurs only at the speed of sound within the material.

Each subsystem is divided into fixed sub-volumes of the computational cell ($12 \times 12 \times 12$ subdivisions in our case) between which the heat transfer occurs. In a typical simulation, part of the excess energy locally introduced by the PKA would be transferred to the electronic subsystem and distributed to the surrounding atoms through electron-phonon coupling. This can heat-up atoms in advance of the pressure wave, but results in lower peaks of the temperature.

In our studies we use the default TTM implementation \cite{fix_ttm_lammps} that also includes electronic stopping, the effect of energy loss of fast moving atoms due to collisions with electrons. However, this implementation assumes constant electronic heat capacity, which we choose to be representative of an electronic temperature of $600 \, \txt{K}$.

To provide a clearer explanation of parameters used in our simulations we will address some details of the two-temperature model as they are given in ref. \cite{duffy_including_2007,rutherford_effect_2007} and \cite{fix_ttm_lammps}. The approach can be refereed to as an inhomogeneous Langevin thermostat
\begin{equation}
  \label{eq:inh_lang}
    m_i \frac{\partial \vec{v}_i}{\partial t} = - \frac{\partial U}{\partial \vec{r}_i}
    - \gamma_i \vec{v}_i
    + \tilde{\vec{F}},
\end{equation}
where $m_i$ is the mass of the $i$-th ion in the system, $\vec{r}$ and $\vec{v}$ are its position and velocity respectively, while $t$ is the time coordinate. The first term on the right-hand side (RHS) represents the force due to the interatomic potential $U$, and the second represents a friction term with coefficient $\gamma_i$. This coefficient can be expressed as $\gamma_i=m_i/D$, where $D$ is a damping factor taken as an input by LAMMPS. This parameter will be particularity important later on as we explain management of the excessive heat introduced by the PKA. The last term is equation \ref{eq:inh_lang} is the random force responsible for the temperature control. The heat transfer is governed by
\begin{equation}
  \label{eq:ttm_heat_trans}
    C_e \rho_e \frac{\partial T_e}{\partial t} = \nabla \cdot (\kappa_e \nabla T_e) - g_p (T_e - T_a) + g_s T_a',
\end{equation}
where $T_e$ and $T_a$ are temperatures of the electronic and the atomic subsystems respectively. Furthermore, $\rho_e$ is the electron density, $C_e$ is the  electronic specific heat and $\kappa_e$ the electronic thermal conductivity. Finally $g_p$ is the coupling parameter for electron-ion interactions while $g_s$ is responsible for the electronic stopping. Parameter $T'_a$, given in units of temperature, corresponds to the kinetic energy of atoms with velocities higher than the cut-off for electronic stopping (equations 7-13 in \cite{duffy_including_2007}). In simulations we define friction coefficients $\gamma_p$ and $\gamma_s$ that are related to $g_p$ and $g_s$ parameters via equations 5 and 6 in ref. \cite{rutherford_effect_2007}. The final parameter used in simulations is the critical velocity $v_\mathrm{c}$ above which electronic stopping is applied.
The values used in simulations are based on literature and are summarised in Table \ref{tab:ttm_prm}. 

\begin{longtable}{lp{2.5cm}lp{2.5cm}lll}

\caption{\label{tab:ttm_prm} Parameters for the TTM used in this work including references that the values are based on. The damping coefficient $D$ is calculated from other parameters assuming a single point TTM model. Critical velocity corresponds to a kinetic energy that is two times the cohesive energy of Zr ($6.469 \mathrm{eV}$). Values are given in both SI and LAMMPS ``metal'' units.}
\\
 & value (SI) & unit (SI) & value ("metal") & unit ("metal") & ref.\\
\hline
\endfirsthead
\multicolumn{6}{l}{Continued from previous page} \\
\hline
 & value (SI) & unit (SI) & value ("metal") & unit ("metal") & ref. \\
\hline
\endhead
\hline\multicolumn{6}{r}{Continued on next page} \\
\endfoot
\endlastfoot
\hline
\(C_e\) & 4.170e-01 & \(\mathrm{J/(K\, mol)}\) & 4.322e-06 & \(\mathrm{eV/K}\) & \cite{jepsen_electronic_1975}\\
\(\rho_e\) & 1.696e+29 & \(\mathrm{m^{-3}}\) & 1.696e-01 & \(\mathring{\mathrm{A}}^{-3}\) & \cite{wang_1994}\\
\(\kappa_e\) & 1.860e+01 & \(\mathrm{J / (K m s)}\) & 1.161e-02 & \(\mathrm{eV} / (\mathring{\mathrm{A}} \mathrm{K} \mathrm{ps})\) & \cite{Crocombette2015}\\
\(\gamma_p\) & 1.797e+11 & \(\mathrm{kg / (mol \, s)}\) & 1.797e+02 & \(\mathrm{g / (mol \, ps)}\) & \cite{wang_1994}\\
$D$ & 5.077e-13 & \(\mathrm{s}\) & 5.077e-01 & \(\mathrm{ps}\) & --\\
\(\gamma_s\) & 3.842e+10 & \(\mathrm{kg} / (\mathrm{mol}\,\mathrm{s})\) & 3.842e+01 & \(\mathrm{g} / (\mathrm{mol}\,\mathrm{ps})\) & \cite{ZIEGLER20101818}\\
\(v_\mathrm{c}\) & 5.231e+03 & \(\mathrm{m} / \mathrm{s}\) & 5.231e+01 & \(\mathring{\mathrm{A}} / \mathrm{ps}\) & --\\
\end{longtable}

\begin{figure}[h!]
\centering
\begin{subfigure}[b]{0.95\textwidth}
\includegraphics[width=1.0\textwidth]{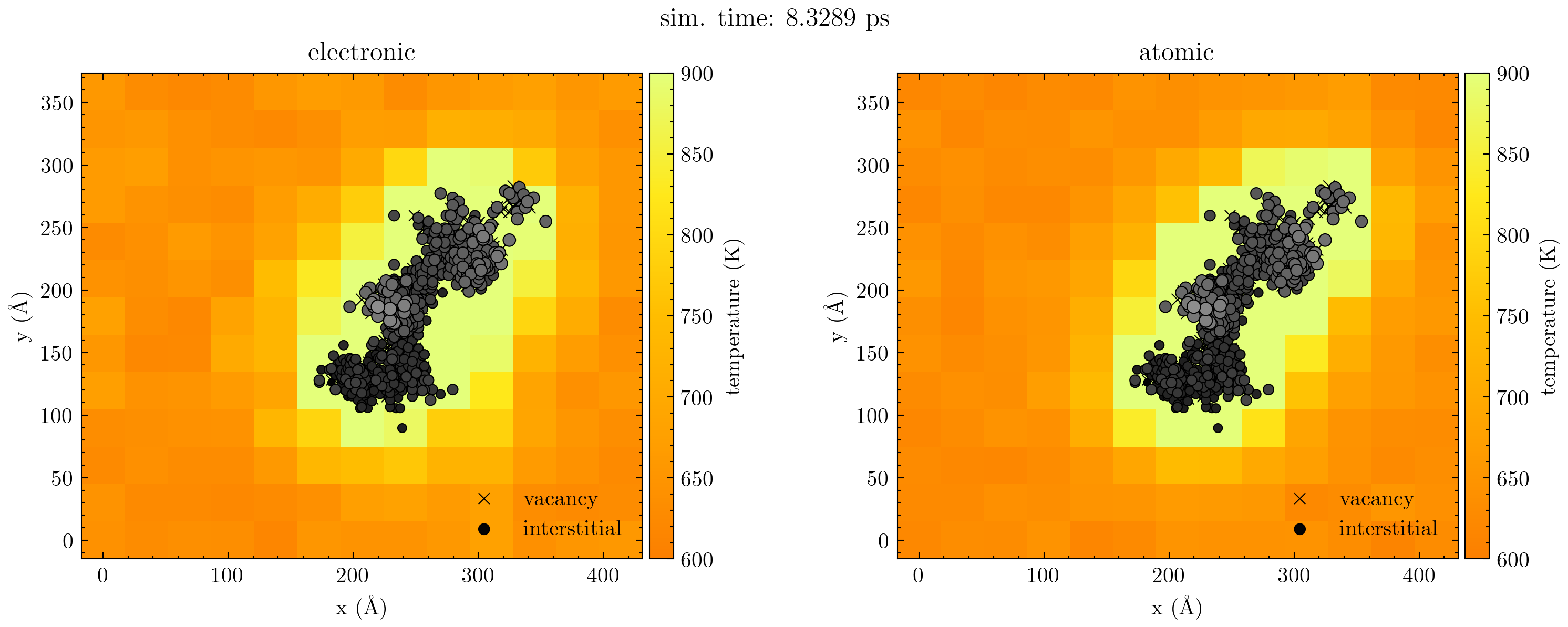}
\caption{\label{fig:org590b21b} Initially concentrated temperature distribution.}
\end{subfigure} 
\begin{subfigure}[b]{0.95\textwidth}
\centering
\includegraphics[width=1.00\textwidth]{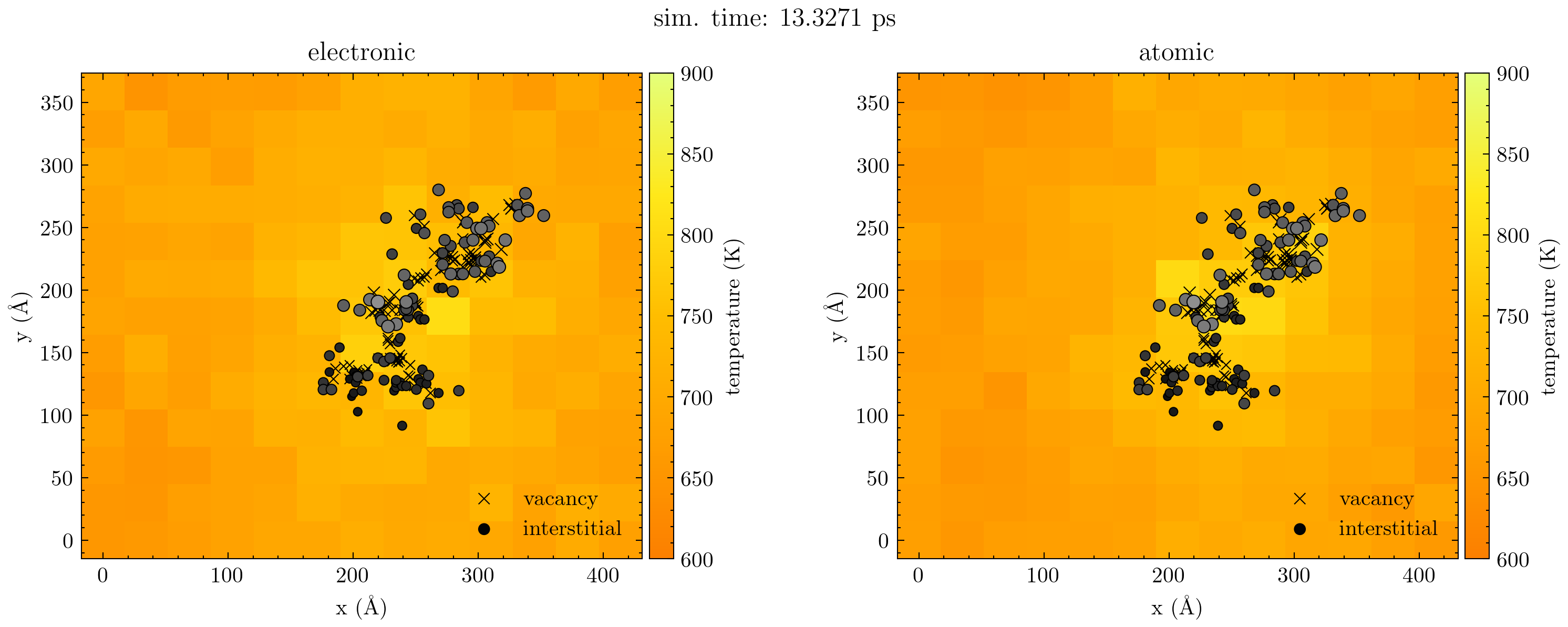}
\caption{\label{fig:orgd135865} Subsequent uniform distribution of the temperature.}
\end{subfigure}
\caption{\label{fig:ttm_dist} The distribution of ``electronic'' and ``atomic'' subsystem temperatures of the TTM during the evolution of the system after PKA was introduced -- (a) thermal spike phase, (b) beginning of recovery. Phases of the collision cascade will be discussed in the next section. What we show is the maximum temperature (marginal) on the axis perpendicular to the plotting coordinates. In other words, given that the distribution is defined as three dimensional matrix, we plot the maximum from columns that otherwise would be obscured by the ``top'' layer. Additionally, we plot the defects estimated using Wigner-Seitz analysis with colours that correspond to the coordinate perpendicular to the screen.}
\end{figure}

The limitation of the TTM, as implemented in LAMMPS, is that it prevents us from using a thermostat that would dissipate the excessive heat introduced by the PKA. This means that recovery would proceed at an elevated temperature (around \(20-30 \, \mathrm{K}\) above $600 \, \txt{K}$ for higher energy PKA's). To avoid this issue we use the fact that after several \(\mathrm{ps}\) the distribution of the electronic temperature is fairly uniform as illustrated in Figure \ref{fig:ttm_dist}.
Since the TTM may be considered a spatially inhomogeneous Langevin thermostat, once we are certain that the temperature distribution is uniform, we can replace it with a homogeneous Langevin thermostat. Hence, beyond this point, we can switch off the TTM and introduce an appropriate thermostat that has a damping coefficient calculated from the TTM parameters (Table \ref{tab:ttm_prm}). However, this can result in overly rapid cooling, since the thermostat implicitly treats the boundaries of our simulations as a perfect heat sink. To avoid this, we introduce a controlled cooling schedule based on an assumption of isotropic heat dissipation from the boundaries of the simulation and the progressive adjustments of the thermostat target temperature. Function that was estimating this temperature was implemented as a Python function that can be run from within the LAMMPS script. The script is an integral part of the associated database.

\begin{figure}[h!]
\centering
  \includegraphics[width=0.5\textwidth]{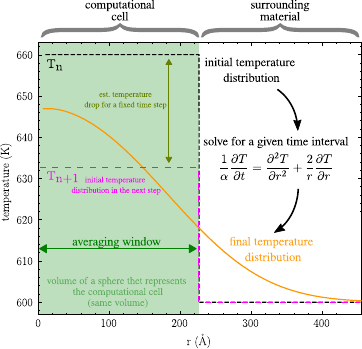}
  \caption{\label{fig:sim_heat_exchange} The concept of estimating the cooling schedule (setting the desired temperature for the thermostat at each sub-iteration). The new target temperature is an average over the surrogate of the computational cell. Each estimates begins with a uniform distribution of temperature at the current temp. of the computational cell. \cmt{probably we need a flow-chart}}
\end{figure}

The process for determining the cooling schedule is as follows. The computational cell, which is assumed to be surrounded by an infinite medium, has excess heat that needs to be dissipated at an appropriate rate. We assume that this cell is well represented by a sphere of equal volume. In spherical coordinates, for an isotropic material, such a problem reduces to solving a one-dimensional heat diffusion equation. Now, consider a time interval for the MD simulation $t_{n} \rightarrow t_{n+1} = t_{n} + \Delta t$. We determine the target temperature $T(t_{n+1})$ for this interval by setting up a temperature distribution in which there is a uniform temperature of $T(t_{n})$ in the region representing the simulation cell. Likewise, the region outside initially has a uniform temperature equal to the final target. We then solve the diffusion equation for this distribution over the period $\Delta t$. The target temperature $T(t_{n+1})$ is given by the average of the evolved temperature distribution over the region representing the simulation cell. The new state of the system becomes a starting point for the next iteration. We repeat the same procedure several times until all excess heat is dissipated. The procedure is illustrated in Figure \ref{fig:sim_heat_exchange}. 

This solution is not perfect, but we believe it improved the realism of our simulations without the necessity of rewriting the TTM module in LAMMPS.

\pagebreak

\section{Cascades simulations -- results and discussion}
\label{sec:cascade_sim}

In this section, we will present the results of the simulations and some key summary statistics that characterise collision cascades.  This will provide a foundation for the generative model (GM) and is a key section for readers interested only in the results of the MD simulations.

\begin{figure}[h!]
\centering
\includegraphics[width=1.00\textwidth]{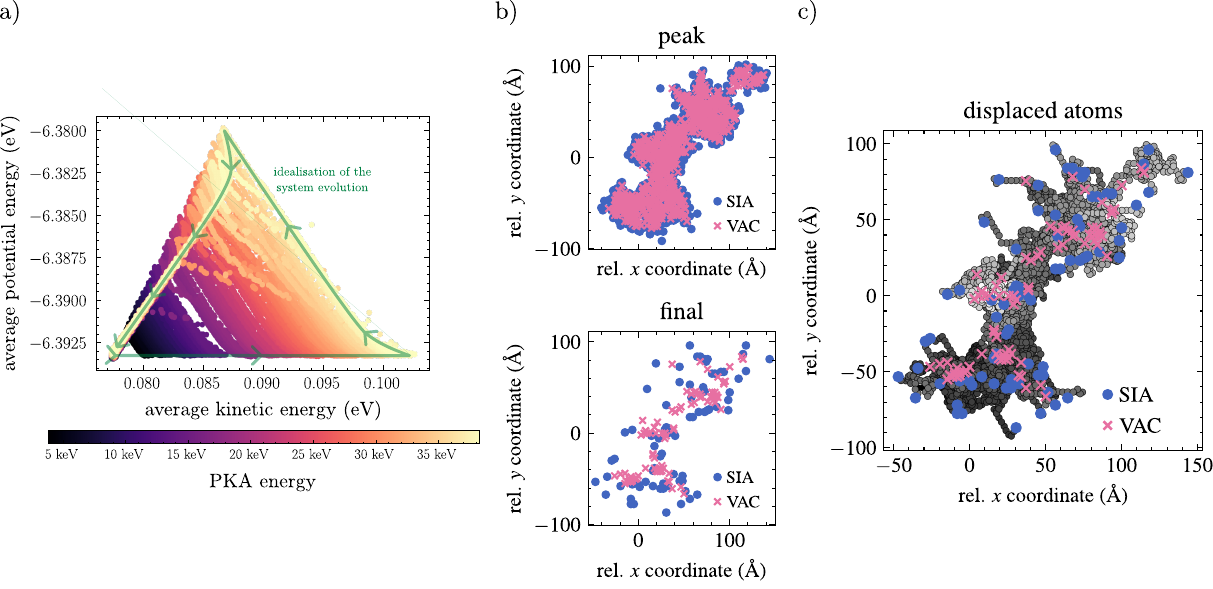}
\caption{\label{fig:cascade_ev} Selected quantities that represent evolution of a $\approx 36 \, \txt{keV}$ collision cascade in Zr with initial temperature $600 \, \txt{K}$. Starting from the left (a), evolution of the average kinetic and potential energy of the system. Green arrows indicate the direction of the evolution. In the middle (b), positions of vacancies (VAC) and self interstitial atoms (SIA), determined  using Wigner-Seitz analysis. Here, we used something we call the ``hot-count'', which means we did not quench the system before calculating the occupancy numbers associated with the equilibrium positions of the crystal. On the right (c), displaced atoms. These atoms changed their nominal positions as a result of a cascade. However, the new position is an equilibrium one in most of the cases. This plot is overlapped with positions of actual stable defects. In this plot shades of grey represent $z$ coordinates. In sub-figures b) and c) coordinates are given in relation to the initial PKA position.}
\end{figure}

In the classical picture given by Bacon \textit{et al.} \cite{bacon_defect_1997} the evolution of collision cascades can be divided into two major phases: ballistic and thermal-spike. A cascade begins with the primary knock-on atom (PKA) colliding with other atoms of the lattice, initiating secondary knock-on atoms, tertiary knock-on atoms and so on. The PKA's kinetic energy is distributed to surrounding atoms displacing them from their equilibrium sites and giving rise to the overall potential energy of the system. This stage can be considered as the ballistic phase. Subsequently, the systems start to evolve towards equilibrium between the potential and kinetic energy of the crystal, resulting in the thermal spike. At this stage, the average kinetic energy can exceed melting temperatures. Subsequently, the heat is dissipated through phonons and electronic transport, and a stable crystal lattice forms again. However, in this process, not all atoms are able to create a perfect lattice. As a result, defects are created, namely pairs of vacancies (VAC) and self-interstitial atoms (SIA). A good visualisation of this process can by found in Figure 1 in \cite{nordlund_improving_2018} and in Figure \ref{fig:cascade_ev} of this work.
Here, we decided to use macroscopic quantities, the average (over atoms) kinetic and the potential energy of the crystal, to illustrate the process. These two coordinates are essentially averaged-out coordinates of the phase space. Figure \ref{fig:cascade_ev} (a) illustrates that the initial spike of the kinetic energy (bottom green line), due to the interaction of PKA with the neutron field, is transformed into the distortion of the crystal, represented by the rise of the system's potential energy (right side of the green triangle). This effect results from subsequent collisions and atomic displacements that accumulate on time scales lower than the thermal vibrations of atoms. The sharp drop in the potential energy indicates the end of the ballistic phase and transition into the thermal spike, followed by the recovery shortly after. It is a more nuanced picture of defect generation than one assumed by Monte-Carlo simulation within the binary collision approximation (BCA). 

Figure \ref{fig:cascade_ev} b) shows the positions of vacancies and self-interstitial atoms at the peak of the cascade and the final configuration of stable defects. Here, we used Wigner-Seitz (WS) analysis to determine nodes (equilibrium atom positions) with zero occupancy (VAC) and greater than one (SIA). Finally, in Figure \ref{fig:cascade_ev} c), we exemplify on high-fidelity data the difference between displaced atoms (or atom replacements) and actual defect production. These quantities are often conflated when employing the NRT-dpa model \cite{nordlund_improving_2018}.

\begin{figure}[h!]
\centering
\includegraphics[width=1.00\textwidth]{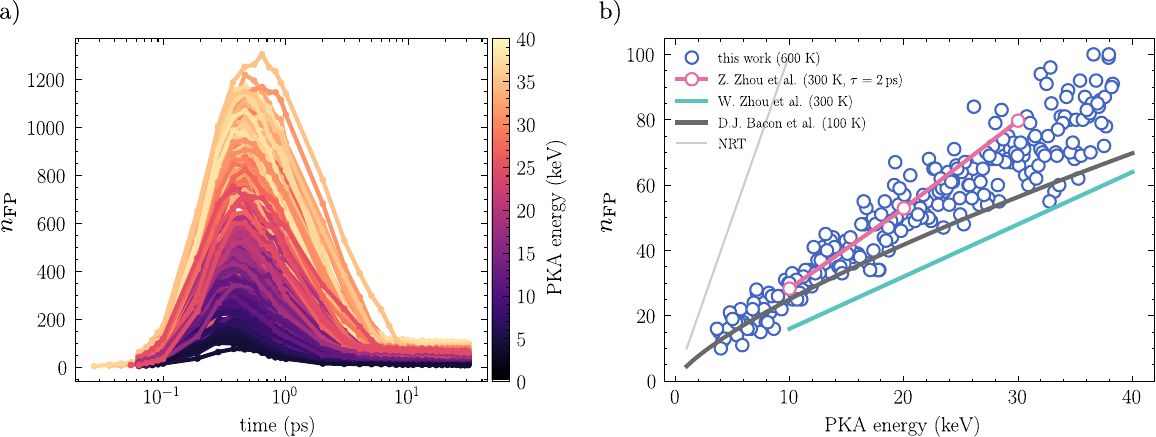}
\caption{\label{fig:dmg_ev_and_compare}Number of Frenkel pairs ($N_{\txt{FP}}$) as estimated using Wigner-Seitz method. On the left (a), evolution during the first $30 \, \txt{ps}$ as a function of time and the PKA energy. On the right (b), comparison across the literature of the number of surviving defects as a function of the PKA energy. The regression lines represent the expected number of defects produced, as evaluated by cited researchers: D. J. Bacon \textit{et al.} -- \cite{bacon_defect_1997}, Z. Zhou \textit{et al.} -- \cite{zhou_molecular_2018}, W. Zhou \textit{et al.} -- \cite{zhou_effects_2020}. For illustration purposes we include the measure of ``dose'' expressed in terms of the NRT model as a number of displacements per atom. In the latter case we assume $40 \, \txt{keV}$ displacement threshold energy. \cmt{We can clearly see that the NRT model can serve as a scalar of the PKA energy at best and has nothing to do with the actual damage of the material. Do we include comment like that?}}
\end{figure}

We now describe key features of the defect distribution resulting from the collision cascades. First, we consider the number of defects produced. Figure \ref{fig:dmg_ev_and_compare} illustrates the count of Frenkel pairs during early stages as well as the number of surviving defects as a function of the PKA energy. 
By comparing values in both sub-figures, again, we clearly see that the relatively small number of surviving defects is a result of a recovery from a highly distorted lattice. However, for the long-term irradiation damage, only the number of surviving defects, presented in Figure \ref{fig:dmg_ev_and_compare} b), matters.

We compare our results with estimates of the expected number of Frenkel pairs (FP) as a function of the PKA energy. When we consider that many simulation parameters can affect the results, e.g. applied potential, effects of temperature, different mechanisms of heat dissipation, presence of the TTM refinement or different values of associated parameters, we can consider that our results are consistent with the literature. This gives us fair confidence in selected potential, parameters controlling the PKA and the thermodynamics of the system.

\begin{figure}[h!]
\centering
\includegraphics[width=1.00\textwidth]{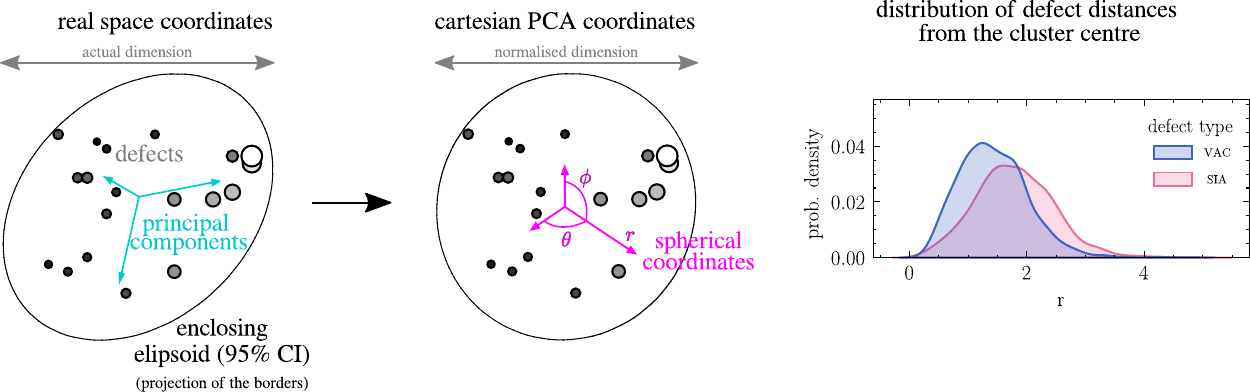}
\caption{\label{fig:desc_vis} Illustration of the description of the defect distribution and size of a defect cluster. From left to right: principal component analysis (PCA) and definition of the size, transformation to normalised, and then to spherical, coordinates, analysis of the distribution. Here, the spherical coordinate is normalised by the standard deviation, not the size of the cascade, of all defects.} 
\end{figure}

From the perspective of damage production and accumulation, the number of defects is only one part of the story. We need to also consider the size of the cluster of defects as well as at least a basic consideration of the distribution. To do this we introduce the following description illustrated in Figure \ref{fig:desc_vis}. Note, that this description is designed to also provide a minimal specification of our generative model which will be further explained in Section \ref{sec:build_cascade_desc}.

The result of each collision cascade, in the form of positions of vacancies (VAC) and self-interstitial atoms (SIA), is considered a cloud of points. To these points, we apply the principal component analysis (PCA), where each one of the components, multiplied by 2, defines an enclosing ellipsoid. The ellipsoid corresponds roughly to the 95\% confidence interval (probability of finding a defect within a volume) and defines the size of a cascade, its shape, as well as position and orientation with respect to the crystal lattice.

At the same time, a new Cartesian coordinate system is defined by the components of the PCA. In our analysis, defect positions are transformed into this system, normalised with respect to the size and shape of the cascade, with associated standard deviation as the unit, and then transformed into spherical coordinates. Transformed points contain information about the relative distance from the cluster centre and concentration in a specific direction (in local PCA coordinates). We quantify these features using the average distance from the centre and parameters of the von Mises-Fisher distribution: concentration $\kappa$ and the expectation $\mu$, \textit{i.e} preferred direction of the angular random variable. In correlation analysis we will use the angle between expectation and the first principal component $\theta_\mu$.

\begin{figure}[h!]
\includegraphics[width=1.00\textwidth]{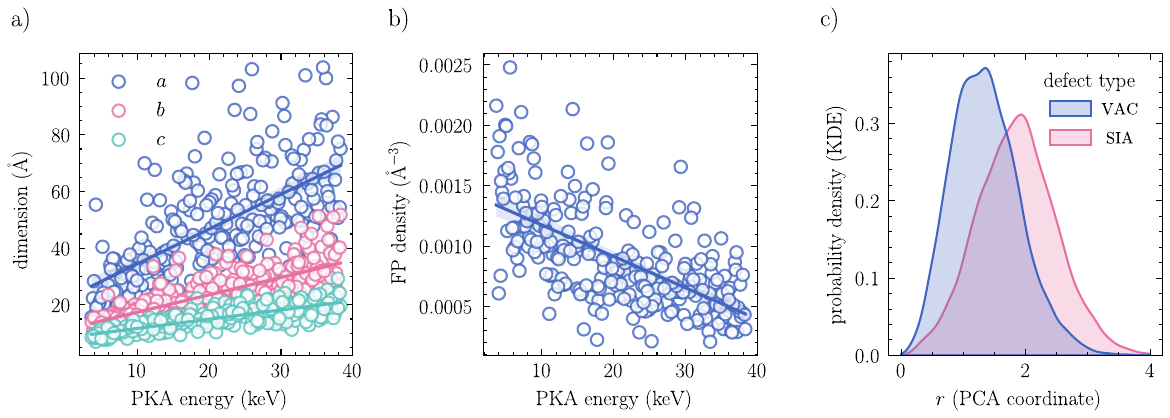}
\caption{\label{fig:desc_res} Dimensions of ellipsoids (ordered with respect to size) $a$, $b$ and $c$ enclosing defect clusters (a), Frenkel-pair density (b), and distribution distances from the cluster centre $r$ within ``local'' PCA coordinates (c). Scatter plots include linear regression with 95\% confidence intervals. Note, that this is not the best estimate of the quantity of interest, only an illustration of the trend. Density is estimated from all data using the Kernel Density Estimation (KDE). Within the PCA coordinates each component quantity is normalised independently by the standard deviations of all defects. Hence, the coverage does not correspond to a sample from a multivariate normal distribution with unit covariance.}
\end{figure}

With the main features of cascades in place, we can have a closer look at some of the relationships. Figure \ref{fig:desc_res} consists of three plots. Starting from the left, in sub-figure (a) we show the dimensions of the cluster. Naturally, the size and particular dimensions increase with the PKA energy. However, the ratio between the principal axes of the enclosing ellipsoid does not correlate with the energy. In other words, we have no evidence that the overall shape of a cascade depends on this quantity. For example, high-energy PKAs do not necessarily result in elongated shapes. Secondly, the increase in size exceeds the increase in the number of defects produced (sub-figure (b)). Therefore, the defect density is lower for higher energy cascades. It is important to remember that the ellipsoid is a very simple representation of shape. As such, branching cascades might not be well represented, and the associated volume may be inaccurately estimated. On the other hand, later analysis suggests that this simplification was sufficient in the sense that in most cases we did not find any irregularities in the estimates. Finally, as expected, vacancies tend to be concentrated closer to the cluster centre than interstitials (sub-figure (c)).

\cmt{We should have used ANOVA here. We should be using ANOVA. In material science, we are way behind.
These considerations are key in the context of damage production and accumulation. However, further quantitative analysis is no longer necessary as it becomes part of the generative model. Now we will focus on more essential qualitative aspects of the results, \textit{i.e.} correlations between the key quantities. We should address these correlations before we begin the design of the model.}

\begin{figure}[!h]
\centering
\includegraphics[width=0.9\textwidth]{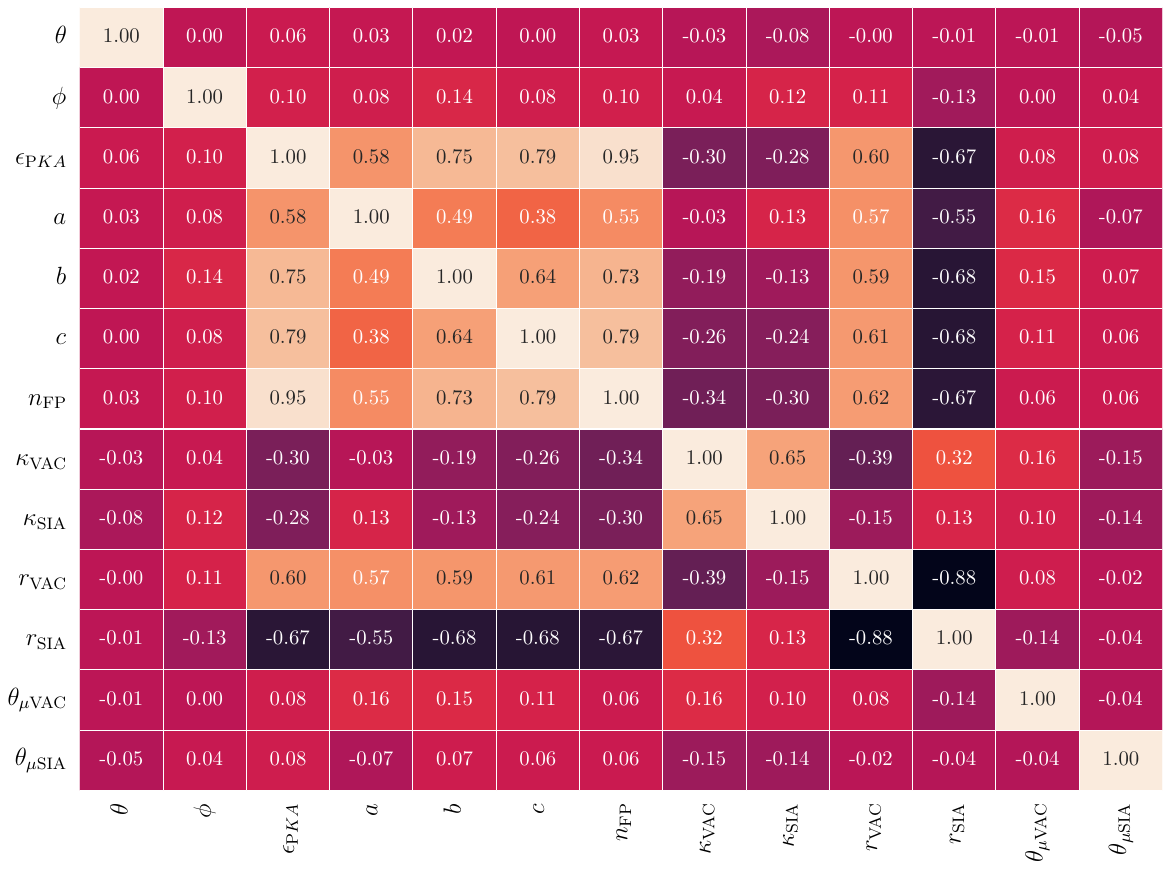}
\caption{\label{fig:corr_heat_map} Symmetric matrix of correlations between key quantities. Here $\phi$ and $\theta$ are the PKA azimuthal and polar angles (in radians), respectively (`$z$ is parallel to the hcp $c$ axis). The PKA energy in eV is represented by $\epsilon_{\mathrm{PKA}}$. Furthermore, the dimensions, starting from the largest, in $\mathring{\text{A}}$ are labelled $a$, $b$ and $c$. The number of Frenkel pairs is $n_{\mathrm{FP}}$. The parameters of the von Mises-Fisher distribution fitted with the maximum likelihood method to the normalised local coordinates are $\vec{\mu}$ (mean direction) and $\kappa$ (concentration parameter). However, here we provide angles ($\theta_{\mu\mathrm{VAC}}$ and $\theta_{\mu\mathrm{SIA}}$) between the first-principle direction of the cluster and the estimated mean direction. These estimates are in local coordinates, so it is essentially an angle between $\vec{\mu}$ and the local $[100]$ direction. Finally, the expected distance of a defect from the cluster centre is given by $r_{\mathrm{VAC}}$ and $r_{\mathrm{SIA}}$.}
\end{figure}

While qualitative assessment can provide indispensable insights into the nature of irradiation damage, we need quantitative analysis to objectivise our conclusions and provide a basis for the development of a regression model. Figure \ref{fig:corr_heat_map} shows the correlations between all the quantities discussed in the above. The coefficients are estimated separately for VAC and SIA. In addition, we include the direction of the PKA momentum, defined by the azimuthal angle $\phi$ and the polar angle $\theta$. These angles are defined by the standard transformation from the global Cartesian coordinates of the lattice (the ``z'' axis is aligned with the ``c'' direction of the hcp crystal) to spherical coordinates. Here, we do not consider the relationship between the direction of the principal component and the direction of the PKA momentum. In our analysis, the former did not correlate with any other quantity of interest. \\
It immediately transpires that the directionality of the PKA is not meaningfully correlated with any other quantity. This can be easily explained. At the temperature of $600 \, \txt{K}$ atoms are significantly displaced from their nominal positions. As a result, there is no well-defined channelling or close-packing direction from a PKA point of view. Therefore, we can ignore the direction of the PKA with respect to the lattice. Furthermore, the expected direction of the defect concentration in the local PCA coordinates, \textit{i.e.} the parameter $\vec{\mu}$ from the von Mises-Fisher distribution, is not correlated with any other quantity either. However, the concentration parameter itself seems to be weakly related to the dimensions of the ellipsoid.

In summary, we have shown that at $600 \mathrm{K}$ we can exclude most of the directional statistics that may be important for cascades occurring at much lower temperatures. A more detailed analysis of these relationships will be undertaken using partial correlation analysis and appropriate regression models, and will form part of the following section where we discuss how to build a hierarchical generative model.


\pagebreak

\section{Generative model}
\label{sec:org9d0697d}

\subsection{Fundamental concepts}
\label{sec:orgbc6212e}

In a broad sense a generative model is one that can ``generate'' representative data. If the model is trained on empirical data, we can use it to interpolate and extrapolate the training set. More formally, we are looking for a joint distribution of causes and outcomes.

In the context of machine learning, generative models are often based on so-called deep-learning and sometimes coupled with a discriminator in adversarial training. However, in this work we will introduce explicit relationships between observable/measurable features. We will try to answer the question what is the most likely position of stable defects after a collision cascade of given energy? We will have to consider a hierarchy of features. As a result, we will also try to pin-point key characteristics (size, shape, number of defects) of the produced defects and provide an empirical relationship with the PKA energy. In other words, we propose to generate directly, instead of from simulation, a representative population of \(n\) defects.

The stating point will be the formal definition of the generative model as a joint density of the cause and effect
\begin{equation}
  p \left( \bm{X}_\mathrm{vac}, \bm{X}_\mathrm{sia}, \vec{x}_{\mathrm{PKA}}, \dot{\vec{x}}_{\mathrm{PKA}} \right) =
  p \left( \bm{X}_\mathrm{vac}, \bm{X}_\mathrm{sia} \mid \vec{x}_{\mathrm{PKA}}, \dot{\vec{x}}_{\mathrm{PKA}} \right)
  \times p \left( \vec{x}_{\mathrm{PKA}}, \dot{\vec{x}}_{\mathrm{PKA}} \right).
\label{eq:gen_model_def}
\end{equation}
In the above, \(\bm{X}_{\mathrm{vac/sia}} \in\mathbb{R}^{n}\times\mathbb{R}^{3}\) are coordinates of defects -- vacancies and self-interstitials respectively, and the density is expanded using the probability chain rule.

We will focus on the first distribution on the right-hand side (RHS), conditioned on the position and momentum of the PKA ($\vec{x}_{\mathrm{PKA}}$,  $\dot{\vec{x}}_{\mathrm{PKA}}$). The prior distribution (second term on the RHS) should represent a specific PKA spectrum. In our simulations it is set to be uniform with respect to the PKA kinetic energy and, as such, it can be considered as a constant. In our derivation the conditional will be expressed in terms of this quantity rather than $\dot\vec{x}_\mathrm{PKA}$.

In general, the result will also depend on the type of the alloy, phase and temperature. However, we are focusing on hcp Zr at \(600 \, \mathrm{K}\) here.

Now we will address a series of simplifying assumptions. This will allow us to express the model in terms that can be explicitly written.

We are using classical molecular dynamics to perform our simulations. The system is deterministic and the result depends only on the generalised positions and momenta of all atoms in the initial configuration. Given the high temperature, different seed during thermalisation, and the exponential instability (Lyapunov) of MD trajectories, we are only ever interested in quantities that do not depend on details of the initial phase space coordinate of the system. Hence, the ``cause'' in the distribution (equation \ref{eq:gen_model_def}) is limited only to the information about the PKA. The complete information implicitly will be reflected in the spread of the final distribution.

Note, that the initial momenta of the PKAs are under our explicit control, and are ``random'' by design (Section \ref{sec:methods}). 

Furthermore, we demonstrated in the previous section that at the temperature of \(600 \, \mathrm{K}\) the direction of the PKA is irrelevant. The key point here is that we can replace the tuple \(( \vec{x}_{\mathrm{PKA}}, \dot{\vec{x}}_{\mathrm{PKA}})\) with the kinetic energy of the PKA -- \(\epsilon_\mathrm{PKA}\).

Up to this point we have been quite casual in writing down the probabilities. As discussed earlier, each simulation starts with a different state and the number of defects will vary from simulation to simulation, even if the initial PKA momentum remains the same. As a result, the event space consists essentially of a collection of vectors (representing positions of defects) of various sizes (alternatively \(n \times 3\) matrices as used before). This prevents us from creating a straightforward statistical model.

Instead, we propose simply to sample position vectors \(n\) times. In such case, correlations between the number of defects and their spatial features will be taken into account through appropriate hierarchical modelling and sequential sampling.

Taking this into account we can write the generative model as a tuple of two distributions
\begin{equation}
p\left(\vec{x}_{\mathrm{vac}},\vec{x}_{\mathrm{sia}}\mid\vec{\beta}\left(\epsilon\right)\right),\,p\left(n\mid\vec{\beta}\left(\epsilon\right)\right),
\label{eq:sampling_useful}
\end{equation}
where \(n\) is the number of defects, and the coordinates of defects \(\vec{x}_\mathrm{vac/sia}\) are dependent on a feature vector \(\vec{\beta}\), which further is a function of the PKA energy \(\epsilon\). 

At this moment the exact features and their numerical representation are not important. We will discuss them in the following subsection. For now, we would like to illustrate that if we follow our logic thoroughly, a key characteristic of our approach emerges almost naturally. Namely, a generative model should be also a hierarchical model.

Consider a simplified example, in which the probability of finding defects is controlled by a set of some attributes \(A\) of a cascade (\(\vec{\beta}\) in \ref{eq:sampling_useful}). This could be e.g. a measure of spread and/or expected distance from the initial PKA position. These attributes can be directly related to the incident (PKA) energy \(E\) as well as the cascade size \(S\), which is set apart from \(A\) in this example. Furthermore, the expected cascade size will also depend on the PKA energy directly. For example, the resulting cascade size can indicate the character of branching and as such, will affect \(A\). Finally, let's assume that all the quantities should be treated as random variables as each result of a simulation can be treated as unique due to thermal fluctuations in atoms' momenta and positions.

In summary, we have three random variables that we need to select before generating positions of the defects, i.e. we need to sample first from a family of distributions. This can be done by sampling from the joint distribution \(p(A,S,E)\) that further can be expanded using the probabilistic chain rule
\begin{equation}
p\left(A,S,E\right)=p\left(A\mid S,E\right)p\left(S,E\right)=p\left(A\mid S,E\right)p\left(S\mid E\right)p\left(E\right).
\label{eq:hierarchy_example}
\end{equation}

The order of expansion is arbitrary in general. However, we can explore the dataset, analyse relationships between \(A\), \(S\) and \(E\) and decide on it. Here we will prefer a more hand-crafted approach rather than analysis of variance common in statistical analysis of experiments. The reason for making the expansion in the first place is to be able to express all correlations as a collection of closed-form parametric estimations.

The final point is that \(p(E)\) is essentially the PKA spectrum, set to be uniform in our dataset, and acts as a normalising factor that does not influence parameters representing the position, shape and scale of other densities. After estimating all of them, we can replace the \(p(E)\) with any PKA spectrum we wish. For example, we can write a routine that introduces in higher-scale simulations a cluster of defects with a realistic probability of being a result of a high or low energy irradiating particle. In the next subsection, we will demonstrate how we can apply those concepts in practice.

\subsection{Building a description of a cascade}
\label{sec:build_cascade_desc}


In this section we will propese a set of features that provide a sufficient description of desired characteristics. In this paper we are trying to keep things as simple as possible. Hence, for now, we relay on heuristics.

The analysis needs to begin with the selection of features we will include in the model. Naturally, we will take into account the number of defects generated. However, we would like to add information about the spatial distribution of defects. It is an essential part as changes in mechanical properties arising from irradiation damage are a result of point defect accumulation and evolution (e.g. assisted diffusion, dislocation loop clusters and void formation). Furthermore, the diffusion happens on much larger time scales than collision cascades. Hence, the initial spread of point defects may assist this evolution in a significant way.

For that reason, the next feature on the list is the size of a cascade. We use the same method as described in the previous section. We employ the principal component analysis (PCA) to the cloud of point defects. Associated principal components are used to estimate an approximately 95\% confidence interval for the probability of finding a defect. This interval is assumed to take the shape of an ellipsoid. Scaled principal components become principal axes of an ellipsoid that define 3 parameters describing the size and shape of a cascade. 

Further refinement of the description is again a reiteration of statistics introduced in Section \ref{sec:cascade_sim}. There, we demonstrated that defects tend to concentrate at certain distances from the initial position of the PKA. Hence, we want to go beyond simple uniform sampling from within the enclosing ellipsoid. In the analysis, we transform the population of defects to the local PCA coordinate system and then to spherical coordinates \((r, \theta, \phi)\). This concept was illustrated in Figure \ref{fig:desc_vis}. The measure of how far from the enclosing ellipsoid defects tend to be will be the average distance from the centre of the ellipsoid. This distance will be calculated separately for vacancies and SIAs. The information about \(\theta\) and \(\phi\) we summarise using parameters of the von Mises-Fisher distribution -- average direction and concentration. 

\subsection{The model hierarchy}
\label{sec:org00cd3f5}

In summary, we propose a description of a collision cascade that consist of: number of Frenkel pairs -- \(n_\mathrm{FP}\), size of the cascade expressed as the length of three principal axes (\(a\), \(b\) and \(c\)) of an ellipsoid that corresponds to the 95\% confidence interval (prob. of finding a defect), the average distance from the cluster centre \(r_\mathrm{sia/vac}\) for self interstitial atoms and vacancies respectively, expected direction of concentration (within transformed PCA coordinates) represented by unit vectors \(\vec{\mu}_\mathrm{vac/sia}\) and concentration \(\kappa_\mathrm{vac/sia}\), both being parameters of the von Mises-Fisher distribution.

In this section, we will investigate relationships between these quantities to establish the hierarchy of the model, as well as the form of the population distribution. In other words, we will decide which characteristics to consider and how to establish a relationship between them. A complete methodology can be summarised in the following steps:
\begin{eqnarray}
  \label{eq:the_path}
  \mathrm{MD\,results} & \rightarrow & \mathrm{description \, vector}  \rightarrow \mathrm{associated \, distributions} \nonumber \\
                       & \rightarrow & \mathrm{"generative" \, distributions} \rightarrow \mathrm{samples\,of\,defects},
\end{eqnarray}
At the end of the process, we want to be able to generate a representative sample of defects that we can test against MD simulations. To do this, we need what we call generative distributions, which are a concept very similar to generative models in machine learning. These distributions will depend on the characteristics of the cascades, which are themselves variable in nature. Therefore, they will also be defined by statistical distributions, which combined will form a distribution of the description/feature vector.

\begin{figure}[h!]
\centering
\includegraphics[width=1.00\textwidth]{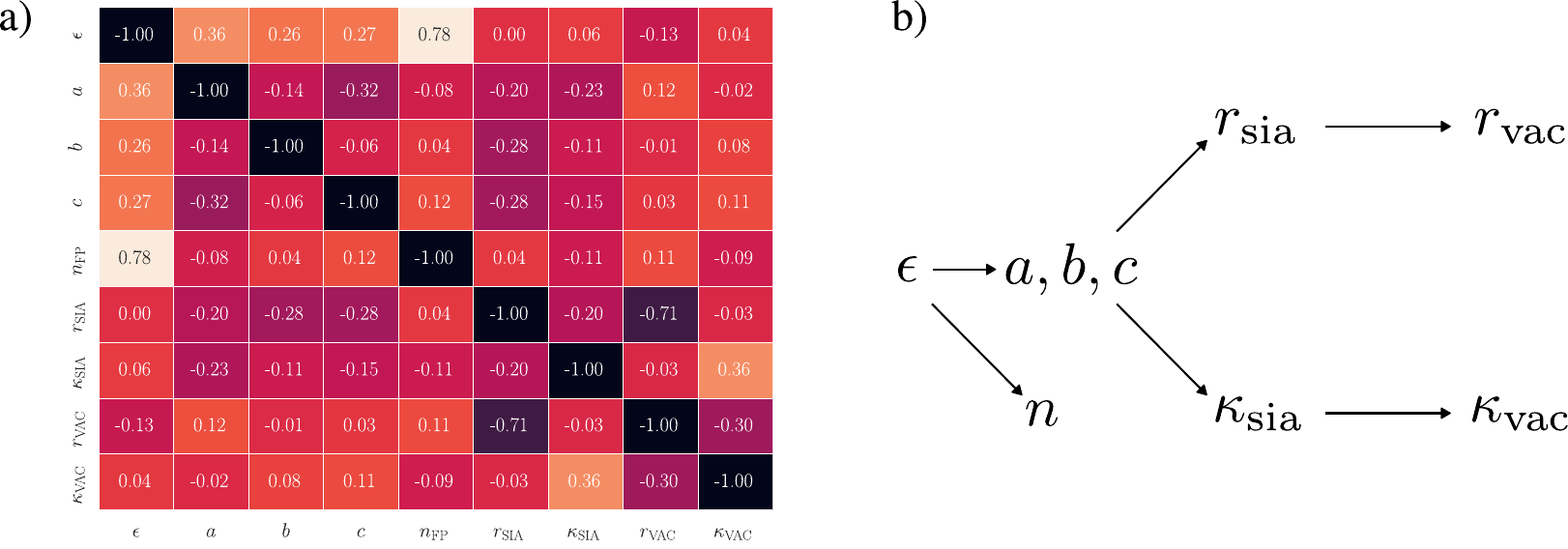}
\caption{\label{fig:part_corr} Sub-figure a) -- Correlation analysis of selected quantities. Sub-figure b) -- model hierarchy based on this information.}
\end{figure}

While correlation might be a sufficient measure (Figure \ref{fig:corr_heat_map}, Section \ref{sec:cascade_sim}) to ignore certain quantities, hierarchical models require a more thorough approach. To avoid issues where two quantities seem to be related to each other, only because they depend on a third, we use partial correlation analysis. Partial correlations measure correlations between residuals resulting from a linear regression. Figure \ref{fig:part_corr} demonstrates this quantity for selected features, as well as a proposed hierarchy, based on the analysis.

As mentioned before, we use a heuristic approach here. For example, we can observe that the size of a cascade (\(a\), \(b\), \(c\)) is related to the PKA energy (\(\epsilon\)) and the expected distance from the defect cluster \(r_\mathrm{sia}\). However, the latter does not correspond to \(\epsilon\). It seems that \(\epsilon\) is related only to the cluster size and the number of Frenkel pairs (\(n\)) -- which is dependent only on \(\epsilon\). Therefore we can infer that in the first level of hierarchy we should have the size (\(a\), \(b\), \(c\)) and population size (\(n\)). Next, we could observe that \(r_\mathrm{sia}\) weakly depends on \(a\), \(b\) and  \(c\) as well as \(r_\mathrm{vac}\). On the other hand, \(r_\mathrm{vac}\) depends mainly on \(r_\mathrm{sia}\). After analysis of dominant partial correlations we found that we can recreate them using the hierarchy presented in Figure \ref{fig:part_corr} b). In our analysis we also used pair plots, omitted here for the sake of brevity, to assess whether the relationship can be approximated by a closed-form solution or whether the correlation is simply a result of incorrectly assigned scales. Finally, since we have a concentration parameter for the von Mises-Fisher distribution, we also need expectations. However, we have discussed earlier that this quantity is not correlated with any other. Hence, it will be sampled as a uniform sampling over the SO3 group (a ``random'' direction).

Next, we will demonstrate how to apply these relationships in a closed form statistical model. In general, the sampling of the feature vector will be done using the joint distribution \(p ( \kappa_\mathrm{vac}, r_\mathrm{vac}, \kappa_\mathrm{sia}, r_\mathrm{sia}, \vec{d}, \epsilon)\) with the following expansion (using probability chain rule)
\begin{eqnarray}
\label{eq:gen_model_full}
p\left(\kappa_{\mathrm{vac}},r_{\mathrm{vac}},\kappa_{\mathrm{sia}},r_{\mathrm{sia}},\vec{d},n,\epsilon\right) & = &
p\left(\kappa_{\mathrm{vac}}\mid r_{\mathrm{vac}},\kappa_{\mathrm{sia}},r_{\mathrm{sia}},\vec{d},\epsilon\right)
p\left(r_{\mathrm{vac}},\kappa_{\mathrm{sia}},r_{\mathrm{sia}},\vec{d},\epsilon\right) \nonumber \\  
& = & \ldots\nonumber \\
& = & p\left(\kappa_{\mathrm{vac}}\mid r_{\mathrm{vac}},\kappa_{\mathrm{sia}},r_{\mathrm{sia}},\vec{d},\epsilon\right) \nonumber \\
& \times & p\left(r_{\mathrm{vac}}\mid\kappa_{\mathrm{sia}},r_{\mathrm{sia}},\vec{d},\epsilon\right) \nonumber \\
& \times & p\left(\kappa_{\mathrm{sia}} \mid r_{\mathrm{sia}},\vec{d},\epsilon\right) \nonumber \\
& \times & p\left(r_{\mathrm{sia}}\mid\vec{d},\epsilon\right) \nonumber \\
& \times & p\left(\vec{d}\mid n,\epsilon\right)p\left(n\mid\epsilon\right)p\left(\epsilon\right),
\end{eqnarray}
where \(\vec{d}\) stands for dimension and \(\vec{d} = (a, b, c)\). As we already mentioned, in general, the order of expansion is arbitrary. We select it on the basis of partial correlations so that we drop most of the conditionals. We might say that (\ref{eq:gen_model_full}) is a fully connected model. If we apply the hierarchy from Figure \ref{fig:part_corr} b), we get
\begin{equation}
\label{eq:final_implicit}
\begin{split}
  & p\left( \kappa_{\mathrm{vac}}, r_{\mathrm{vac}}, \kappa_{\mathrm{sia}}, r_{\mathrm{sia}}, \vec{d}, \epsilon \right) = \\ 
  & p\left( \kappa_{\mathrm{vac}}\mid\kappa_{\mathrm{sia}}\right)p\left(r_{\mathrm{vac}}\mid r_{\mathrm{sia}} \right)
  p\left( \kappa_{\mathrm{sia}}\mid\vec{d},\epsilon\right)p\left(r_{\mathrm{sia}}\mid\vec{d},\epsilon \right)
  p\left( \vec{d}\mid\epsilon\right) p\left(n\mid\epsilon\right) p\left( \epsilon \right).
\end{split}
\end{equation}
Another way to look at this is that we take an arbitrary expansion using the probability chain rule and remove from conditionals those quantities that do not correlate. In the process, we select an expansion that is simplest and informed by the associated physics.

\subsection{Explicit form of distributions}
\label{sec:explicit_dist}

One of the best features of using the expansion \ref{eq:final_implicit} is the fact that we can express a complex multivariate distribution as a product of simpler ones. While our approach is somewhat related to building sophisticated Bayesian hierarchical models, we adopt a simpler view, in the hope that this will make the model more accessible to the wider community and lay foundations for developing phenomenological relationships between features of cascades. 

In general, we regard the problem as a sequence of non-linear regression sub-problems. However, with the right data transformation we will be able to linearise them and also assume a constant variance/covariance of the transformed data. Furthermore, we will make a simplifying assumption that the variability is normally distributed in the transformed coordinates. As a result, most of the time the expectation can be simply estimated with ordinary least squares using
\begin{equation}
\boldsymbol{w}=\left( \boldsymbol{X}^{\top} \boldsymbol{X} \right)^{-1} \boldsymbol{X}^{\top}\mathbf{y},
\end{equation}
where \(X\) is the design matrix and data-values \(y\) are approximated by \(y=\boldsymbol{X}\boldsymbol{w}\). Results of the regression will serve as a model of expectation in the distribution we will assume for our model. Likewise, the variance will be estimated using the estimator
\begin{equation}
s^{2}={\frac {{\hat {\varepsilon }}^{\top}{\hat {\varepsilon }}}{n-p}}
\end{equation}
where \(\hat\epsilon\) is the vector of residuals, \(n\) is number of data-points and \(p\) is the dimensionality of the data.

This brings us to the final step of more general considerations -- selecting a parametric family for each distribution. Since the process is extensive, we will follow steps thoroughly only at the beginning of the hierarchy. The goal here is to introduce all the concepts that are necessary to recreate the results. The complete set will be summarised at the end of the section in table \ref{tab:h_model}. We also would like to reiterate some basic concepts from statistics and give this section some characteristics of a tutorial.   

Selecting the distribution for \(\epsilon\) (PKA energy) is trivial, as it was set to be uniform in the data-set. Hence, \(p\left(\epsilon\right)\propto\mathrm{const.}\). This distribution can be adjusted according to need. However, we are focusing on the estimation of all parameters based on generated data.

It is well established that the number of Frenkel-Pairs can be represented by a power-law \cite{bacon_computer_1995},
\begin{equation}
n\left(\epsilon\right)=w_{n}\epsilon^{k_{n}}.
\end{equation}
Hence, it is a linear relationship in the log-log domain. We make a leap of faith and assume that in this domain, the variance is also well represented by the normal distribution. This is a common assumption in the linear regression. Then the (probability of generating a data-point will have the form
\begin{equation}
p\left(\ln n\mid\ln\epsilon\right)=\frac{1}{\sqrt{2\pi\sigma_{n}^{2}}}\exp\left(-\frac{\left(\ln n-\mu_{\ln n\mid\ln\epsilon}\right)^{2}}{2\sigma_{n}^{2}}\right),
\end{equation}
where the expectation is given by
\begin{equation}
\label{eq:org5d7e4dd}
\mu_{\ln n\mid\ln\epsilon}=\ln\left(w_{n}\epsilon^{k_{n}}\right)=k_{n}\ln\epsilon+\ln w_{n}.
\end{equation}
In the above, we explicitly wrote the argument as a logarithm of the quantity of interest, but it should be here regarded as a variable in transformed coordinates.

Here we can see that the transformation introduces a linear relationship between the logarithm of PKA energy and the logarithm of the number of defects generated. To evaluate the parameters of this relationship we will use the least squares method. This is equivalent to selecting a maximum likelihood estimate and assuming uniform Gaussian ``noise''. During the sampling, we will use these estimates as parameters of the distribution. Such a procedure corresponds to sampling from a posterior predictive with a prior distribution set as a Dirac delta function. In other words, we make a simplifying assumption that our estimates are exact, i.e. there is no associated uncertainty. This theme will be recurrent throughout the procedure. 

\begin{figure}[h!]
\centering
\includegraphics[width=1.00\textwidth]{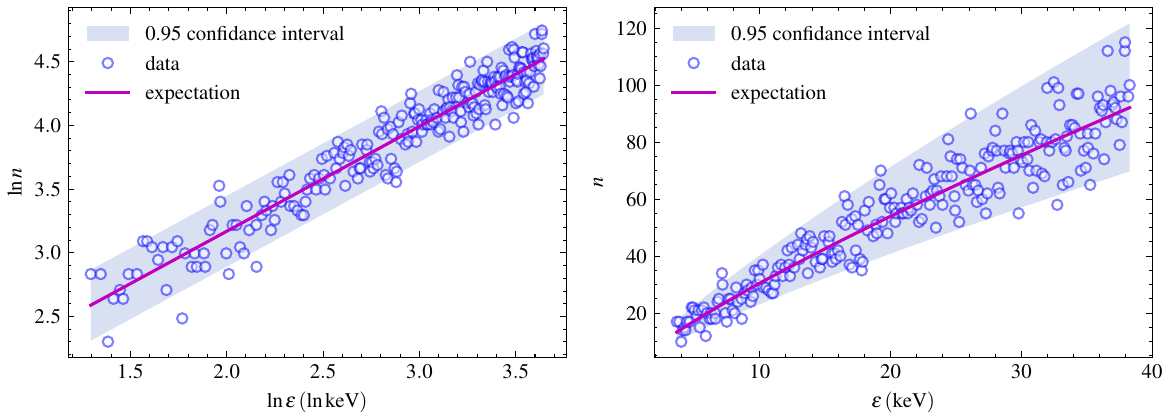}
\caption{\label{fig:model_fit_n} The number of defects produced in a collision cascade as a function of the PKA energy. Figures include the results of MD simulations (the data), ``predicted'' expected number of defects and approximated 95\% confidence interval. In the latter case we used the 1.96 expansion coefficient. Note, that the confidence interval represents the spread of the data very well across the whole domain of the data.}
\end{figure}

Nevertheless, we are interested in the distribution \(p\left(n\mid\epsilon\right)\). Although in practice, we will generate data in the log-log domain and transform them accordingly. Consider a textbook example of two random variables \(X\) and \(Y\). Let \(Y=f\left(X\right)\), and \(f\) is a bijective (as it is in this case, a ``complete'' one-to-one map). Then 
\begin{equation}
p_{Y}\left( y \right)= p_{X} \left( f^{-1} \left( y \right) \right) \left| \frac{ df^{-1} \left( y \right) }{dy} \right|.
\end{equation}
This is a well known example of the change of variables formula. In this form it is applicable to monotonic functions. It is also appropriate to conditional probabilities. For example, If we assume that the relationship between the number of FP pairs $n$ and the PKA energy $\epsilon$ is well approximated by a power law, then from the regression we can easily estimate expectation of $\ln n$. As we are interested in the distribution of $n$, we take \(f \left( \cdot \right) = \exp \left( \cdot \right) \implies f^{-1} \left( \cdot \right) = \ln \left( \cdot \right)\). Therefore
\begin{equation}
p\left(n\mid\epsilon\right)=\frac{1}{\sqrt{2\pi\sigma_{n}^{2}}}\exp\left(-\frac{\left(\ln n-\mu_{\ln n\mid\ln\epsilon}\right)^{2}}{2\sigma_{n}^{2}}\right)\times\left|\frac{1}{n}\right|.
\end{equation}
It is no surprise that we obtained the log-normal distribution. This distribution will represent the variability in the number of Frenkel pairs with the given PKA energy. In the above, the variance is a single parameter, which is an assumption of the transformation. Having established the distribution we can assess the quality of our assumptions by examining the distribution and the data. The results are presented in Figure \ref{fig:model_fit_n}.
It immediately transpires that the power law is not only a great way to represent the expectations, but also that the predicted confidence interval corresponds to the spread of data very well. 

\begin{figure}[h!]
\centering
\includegraphics[width=1.00\textwidth]{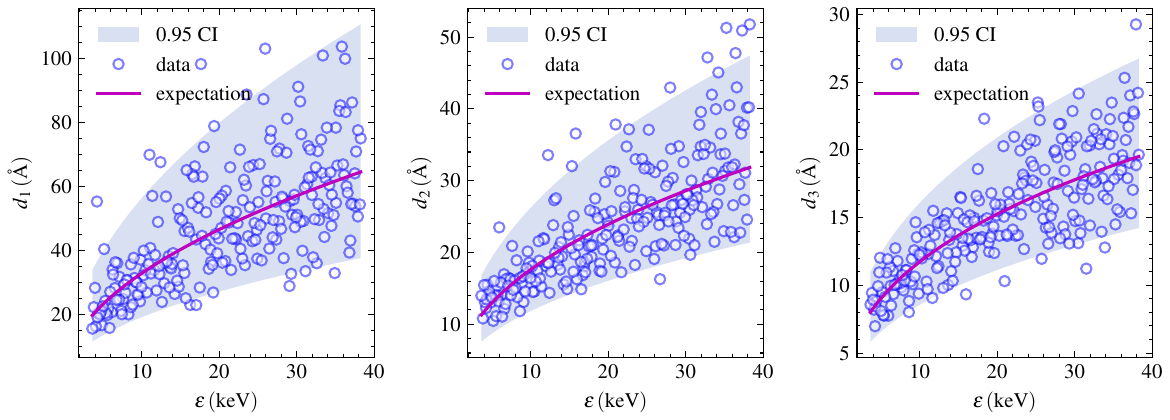}
\caption{\label{fig:model_fit_size} Simulated and predicted sizes (defined via an enclosing ellipsoid) of collision cascades as a function of the PKA energy. The primary focus here is on how well the distribution represents the data. Hence, axes scales are chosen with visibility in mind. However, note that the first and largest principal axis is significantly larger. CI refers to the confidence interval. \cmt{I would like to keep it this way to show how well the model shapes the data}}
\end{figure}

According to our earlier propositions, at the same level of hierarchy we have the size of a cascade. This relationship is a bit more complicated as the size is represented by a vector \( \vec{d} = (a, b, c) \) or $\left(d_1, d_2, d_3 \right)$. In this case we will also take advantage of the flexibility of the power law and assume that each component \( d_i \) of the vector \(\vec{d}\) is given by
\begin{equation}
d_{i}=w_{d_{i}}\epsilon^{k_{d_{i}}}\implies\ln d_{i}=k_{d_{i}}\ln\epsilon+\ln w_{d_{i}}.
\end{equation}
Regarding the distribution we follow a similar reasoning as before, but this time we will us a multivariate Gaussian distribution
\begin{equation}
p\left(\ln \vec{d} \mid \ln \epsilon \right)=\left(\left(2\pi\right)^{3}\left|\bm{\sigma}\right|\right)^{-1/2}\exp\left(-\frac{1}{2}\left(\ln\vec{d}-\vec{\mu}_{\ln d}\right)^{\top}\bm{\sigma}^{-1}\left(\ln\vec{d}-\vec{\mu}_{\ln d}\right)\right),
\end{equation}
where \(\left|\cdot\right|\) is a determinant. For many variables the change of variables is factored by a Jacobian \(\bm{J}\), i.e.
\begin{equation}
\label{eq:inv_trans}
p_{Y}\left(y\right)=p_{X}\left(\vec{f}^{-1}\left(y\right)\right)\left|\bm{J}\right|,
\end{equation}
where
\begin{equation}
J_{ij}=\frac{\partial f_{i}^{-1}\left(y\right)}{\partial x_{j}}.
\end{equation}
The transformation \(\vec{f}\left(\vec{x}\right)\) and associated Jacobian are defined as
\begin{equation}
f_{i}\left(\vec{x}\right)=\exp\left(x_{i}\right)\implies J_{ij}=\frac{\partial f_{i}^{-1}\left(\vec{d}\,\right)}{\partial d_{j}}=\begin{cases}
1/d_{i} & i=j\\
0 & i\neq j
\end{cases}.
\end{equation}
Finally the transformed density will take form 
\begin{eqnarray}
\label{eq:dist_size}
p\left(\vec{d}\mid\epsilon\right) 
& = & \left(\left(2\pi\right)^{k}\left|\bm{\sigma}\right|\right)^{-1/2}
\left(\prod_{i=1}^{3}\frac{1}{d_{i}}\right) \nonumber \\
& \times &
\exp\left(-\frac{1}{2}\left(\ln\vec{d}-\vec{\mu}_{\ln d}\right)^{\top}\bm{\sigma}^{-1}\left(\ln\vec{d}-\vec{\mu}_{\ln d}\right)\right).
\end{eqnarray}
Results in Figure \ref{fig:model_fit_size} illustrate that our statistical model represents data from MD simulation quite well.
Furthermore, we can see that the proportions between principal axes, at least in terms of expectations, are preserved.

Subsequently, we would like to make the estimates for \(r_\mathrm{sia}\), namely the average distance of a self interstitial atom from the cluster centre in local PCA coordinates, as a function of the cluster dimensions. This time we will skip an explicit derivation of densities as the procedure will be quite similar. However, the model for the data will require more explanation. To be able to use a simple transformation that will linearise the relationship we need to introduce an additional bias. This time the power law is not an appropriate model. Instead we will use a simple exponential function of a form
\begin{equation}
r_{\mathrm{sia}}=\exp\left(-\vec{w}_{\mathrm{sia}}\cdot\vec{d}+b_{\mathrm{d}}\right)+r_{\infty}\implies\ln\left(r_{\mathrm{sia}}-r_{\infty}\right)=-\vec{w}_{\mathrm{sia}}\cdot\vec{d}+b_{\mathrm{d}}'
\end{equation}
where we have introduced a baseline parameter \(r_\infty\). This quantity can be regarded as the average radius in the limit of an infinitely large cascade. In the course of our analysis, we noticed that  \(r_\mathrm{vac}\) and \(r_\mathrm{sia}\) seem to converge to this value. In the first case ($\vec{r}_\mathrm{vac}$) it approaches the value from below, and in the other ($\vec{r}_\mathrm{sia}$) from above as it decreases. Note, that we are speaking here about relative/normalised PCA coordinates. To estimate \(r_\infty\) we simply fitted two exponential functions (with a negative exponent) at the same time assuming that they share the same bias (value at the infinity). Having done that, we used this result as a fixed value of \(r_\infty\). This allowed us to calculate the differences \( r_{\mathrm{sia/vac}} - r_\infty \) from (\ref{fig:model_fit_size}), before applying the logarithmic transformation and estimation of  \(\vec{w}_\mathrm{sia}\) and \(b_\mathrm{d}\).

The resulting distribution again will be the univariate log-normal, as we made the same assumption about the normality of the noise. The difference here is that the linear, or rather linearised, model for the expectation will not depend on the logarithmic transformation only on the direct scalar product of the size \(\vec{d}\), associated weights \(\vec{w}_\mathrm{sia}\) and bias \(b_\mathrm{d}\).

\begin{figure}[h!]
\centering
\includegraphics[width=1.00\textwidth]{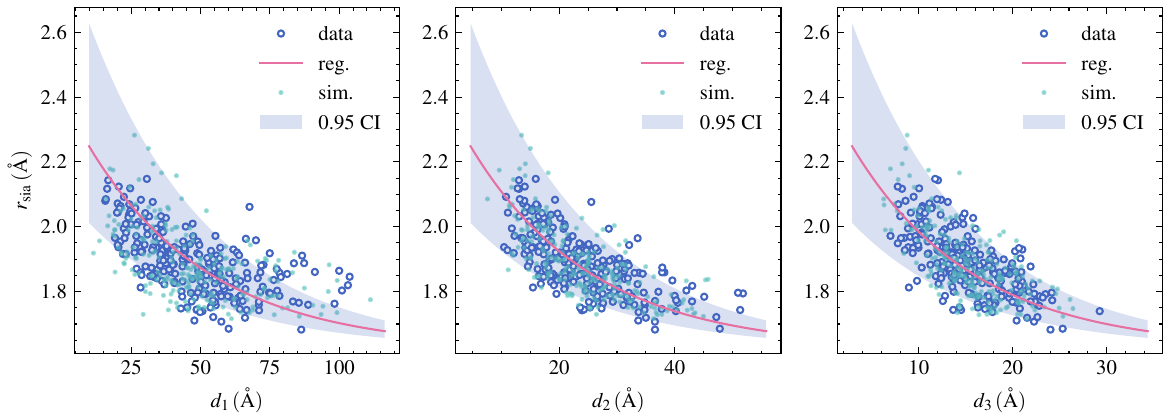}
\caption{\label{fig:r_sia_est} Regression of the average distance of self-interstitial atoms (SIA) as a function of the cascade dimensions. The regression line (reg.) is in four-dimensional space of $\vec{d} \times r_{\mathrm{sia}}$. The expectation, bounds of confidence interval (CI) (scaled $2\times$ variance) and the data are plotted as a planar projection on a given plane. The resampled/simulated data (sim.) illustrate a successful regression.}
\end{figure}


The results of this estimation (in the case of SIA) are illustrated in Figure \ref{fig:r_sia_est}. As we move up in the hierarchy of the model, we need to consider that some variations in the observed data are a consequence of variations in multidimensional arguments. For this reason, a simple visual assessment of the confidence interval might be insufficient. Our approach is to assess the goodness of fit by simply plotting resampled data. However, the procedure is not straightforward anymore. More information will be provided in Section \ref{sec:reg_results}. For now, it is sufficient to say that we are using the Markov chain Monte Carlo method and the PyMC3 python package \cite{Salvatier2016}. Each quantity of interest, from each stage of the hierarchy, is sampled subsequently, and each time it defines the distributions in the following stage.

Further derivations of features of our model (Figure \ref{fig:part_corr} b) are deduced using very similar logic and do not require additional comment. Therefore, we will omit explicit illustrations of further derivations in order to provide a more concise narrative. Needless to say, the performance of the resampling is similar in other cases.

A complete list of all components of the hierarchical model \ref{eq:gen_model_full} is summarised in Table \ref{tab:h_model}.
\begin{table}
\begin{center}
\scalebox{0.75}{
\begin{tabular}{ccccccc}
 & i.v. & q.o.i. & model & $\mu$ / $\vec{\mu}$ & $f^{-1}\left(\cdot\right)$ & $p\left(\cdot\right)$\tabularnewline
\hline 
I & $\epsilon$ & $n$ & $w_{n}\epsilon^{k_{n}}$ & $k_{n}\ln\epsilon+\ln w_{n}$ & $\ln\left(\cdot\right)$ & $\Phi\left(\ln n,\mu,\sigma_n\right)\frac{1}{n}$\tabularnewline
I & $\epsilon$ & $\vec{d}$ & $w_{d_{i}}\epsilon^{k_{d_{i}}}$ & $\vec{k}_{\vec{d}}\ln\epsilon+\ln\vec{w}_{\vec{d}}$ & $\ln\left(\cdot\right)$ & $\Phi_{\mathrm{MV}}\left(\ln\vec{d},\vec{\mu},\boldsymbol{\sigma}^{\left(\vec{d}\right)}\right)\left(\prod\limits _{i=1}^{3}\frac{1}{d_{i}}\right)$\tabularnewline
II & $\vec{d}$ & $r_{\mathrm{sia}}-r_{\infty}$ & $\exp\left(-\vec{w}_{r_{\mathrm{sia}}}\cdot\vec{d}+b_{r_{\mathrm{sia}}}\right)$ & $-\vec{w}_{r_{\mathrm{sia}}}\cdot\vec{d}+b_{r_{\mathrm{sia}}}$ & $\ln\left(\cdot\right)$ & $\Phi\left(\ln\left(r_{\mathrm{sia}}-r_{\infty}\right),\mu,\sigma_{r_\mathrm{sia}}\right)\frac{1}{r_{\mathrm{sia}}-r_{\infty}}$\tabularnewline
II & $\vec{d}$ & $\kappa_{\mathrm{sia}}$ & $\exp\left(-\vec{w}_{\kappa_{\mathrm{sia}}}\cdot\vec{d}+b_{\kappa_{\mathrm{sia}}}\right)$ & $-\vec{w}_{\kappa_{\mathrm{sia}}}\cdot\vec{d}+b_{\kappa_{\mathrm{sia}}}$ & $\ln\left(\cdot\right)$ & $\Phi\left(\ln\kappa_{\mathrm{sia}},\mu,\sigma_{\kappa_{\mathrm{sia}}}\right)\frac{1}{\kappa_{\mathrm{sia}}}$\tabularnewline
III & $r_{\mathrm{sia}}$ & $r_{\mathrm{vac}}$ & $\left(w_{r_{\mathrm{vac}}}r_{\mathrm{sia}}+b_{r_{\mathrm{vac}}}\right)^{1/3}$ & $w_{r_{\mathrm{vac}}}r_{\mathrm{sia}}+b_{r_{\mathrm{vac}}}$ & $\left(\cdot\right)^{3}$ & $\Phi\left(r_{\mathrm{sia}},\mu,\sigma_{r_{\mathrm{vac}}}\right)3r_{\mathrm{sia}}^{2}$\tabularnewline
III & $\kappa_{\mathrm{sia}}$ & $\kappa_{\mathrm{vac}}$ & $\exp\left(w_{\kappa_{\mathrm{sia}}}\kappa_{\mathrm{vac}}+b_{\kappa_{\mathrm{sia}}}\right)$ & $w_{\kappa_{\mathrm{sia}}}\kappa_{\mathrm{vac}}+b_{\kappa_{\mathrm{sia}}}$ & $\ln\left(\cdot\right)$ & $\Phi\left(\kappa_{\mathrm{sia}},\mu,\sigma_{\kappa_{\mathrm{vac}}}\right)\frac{1}{\kappa_{\mathrm{sia}}}$\tabularnewline
\end{tabular}
}
\end{center}
\caption{The list of distributions for the hierarchical model defined in equation (\ref{eq:gen_model_full}). The first column indicates the layer in the hierarchy (figure \ref{fig:part_corr}). The independent variable (i.v.) from one layer is the quantity of interest (q.o.i.) of the previous one. Column ``model'' refers to the equations that define the relationships, while $\mu$ is the associated model of the expectation. We also provide the inverse transformation $f^{-1}$ that is used to derive (via equation \ref{eq:inv_trans}) the form of the distribution in the last column. The function $\Phi$ represents the normal distribution, while the label MV indicates that in fact, it is the multivariate normal distribution. Dummy variable $\mu$ is a placeholder for what comes out of the model}
\label{tab:h_model}
\end{table}
\pagebreak

\subsection{Results of the regression and evaluation of the sampling procedure}
\label{sec:reg_results}
While previously we defined all components for the hierarchical model, here we will address details of the sampling procedure, present estimated values of all parameters and asses its usefulness.

We emphasized deriving an explicit form for the distributions of cascade features. However, in practice, all the features of the cascades are sampled in the transformed coordinates. This way, we can limit ourselves to using Gaussian distributions most of the time. For example, we would sample $\log n_{\mathrm{FP}}$ from the Gaussian distribution rather than $n_{\mathrm{FP}}$ from the log-normal.

\begin{figure}[h!]
\centering
\includegraphics[width=0.75\textwidth]{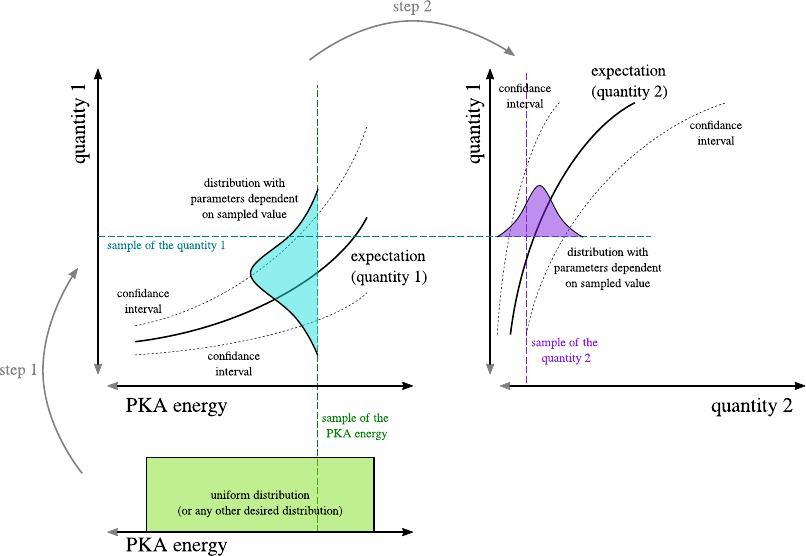}
\caption{\label{fig:model_sampling_concept} The concept of sampling in hierarchical modelling of collision cascades.}
\end{figure}

In practice we use the standard Model class from PyMC3 library to define the distribution and sample using the Markov Chain Monte Carlo method. The logical order of the procedure can be described as follows. After sampling a feature (\textit{e.g.} PKA energy), we apply an appropriate transformation (using a deterministic function $f$ that can be inferred from Table \ref{tab:h_model}) and proceed further to the next stage of the hierarchy. There, we use this transformed sample to evaluate the expectation of the quantity from the next stage of the hierarchy. We continue this process until all features have been sampled. The concept is illustrated in Figure \ref{fig:model_sampling_concept}.

All the parameters necessary to evaluate distributions in the hierarchy (equation \ref{eq:gen_model_full} and Table \ref{tab:h_model}) are presented in Table \ref{tab:h_model_num}. We have estimated them using the procedure described in Section \ref{sec:explicit_dist}.
\begin{table}
\begin{center}
\scalebox{0.75}{
\begin{tabular}{cccccccccc}
quantity &  & \multicolumn{3}{c}{expectation} &  & \multicolumn{4}{c}{standard deviation / covariance}\tabularnewline
\noalign{\vskip\doublerulesep}
\hline 
\multirow{2}{*}{$n$} &  & $\ln w_{n}$ & $k_{n}$ &  &  & $\sigma_{n}$ &  &  & \tabularnewline
\cline{3-4} \cline{4-4} \cline{7-7} 
\noalign{\vskip\doublerulesep}
 &  & 1.523 & 0.823 &  &  & 0.142 &  &  & \tabularnewline
\hline 
\noalign{\vskip\doublerulesep}
\multirow{4}{*}{$\vec{d}$} &  & $\ln w_{d_{1}}$ & $\ln w_{d_{2}}$ & $\ln w_{d_{3}}$ &  & $\boldsymbol{\sigma}^{\left(\vec{d}\,\right)}$ & $\sigma_{:1}^{\left(\vec{d}\,\right)}$ & $\sigma_{:2}^{\left(\vec{d}\,\right)}$ & $\sigma_{:3}^{\left(\vec{d}\,\right)}$\tabularnewline
\cline{2-5} \cline{3-5} \cline{4-5} \cline{5-5} \cline{7-10} \cline{8-10} \cline{9-10} \cline{10-10} 
\noalign{\vskip\doublerulesep}
 &  & 2.334 & 1.853 & 1.590 &  & $\sigma_{1:}^{\left(\vec{d}\,\right)}$ & 0.076 & 0.000 & -0.009\tabularnewline
\cline{2-5} \cline{3-5} \cline{4-5} \cline{5-5} \cline{7-7} 
\noalign{\vskip\doublerulesep}
 &  & $k_{d_{1}}$ & $k_{d_{2}}$ & $k_{d_{3}}$ &  & $\sigma_{2:}^{\left(\vec{d}\,\right)}$ & 0.000 & 0.041 & 0.005\tabularnewline
\cline{2-5} \cline{3-5} \cline{4-5} \cline{5-5} \cline{7-7} 
\noalign{\vskip\doublerulesep}
 &  & 0.503 & 0.441 & 0.379 &  & $\sigma_{3:}^{\left(\vec{d}\,\right)}$ & -0.009 & 0.005 & 0.026\tabularnewline
\hline 
\noalign{\vskip\doublerulesep}
\multirow{2}{*}{$r_{\infty}$} &  & $r_{\infty}$ &  &  &  &  &  &  & \tabularnewline
\cline{2-3} \cline{3-3} 
\noalign{\vskip\doublerulesep}
 &  & 1.620 &  &  &  &  &  &  & \tabularnewline
\hline 
\noalign{\vskip\doublerulesep}
\multirow{4}{*}{$r_{\mathrm{sia}}$} &  & $w_{r_{\mathrm{sia}},1}$ & $w_{r_{\mathrm{sia}},2}$ & $w_{r_{\mathrm{sia}},3}$ &  & $\sigma_{r_{\mathrm{sia}}}$ &  &  & \tabularnewline
\cline{3-5} \cline{4-5} \cline{5-5} \cline{7-7} 
\noalign{\vskip\doublerulesep}
 &  & -0.005 & -0.016 & -0.034 &  & 0.242 &  &  & \tabularnewline
\noalign{\vskip\doublerulesep}
 &  & $b_{r_{\mathrm{sia}}}$ &  &  &  &  &  &  & \tabularnewline
\cline{3-3} 
\noalign{\vskip\doublerulesep}
 &  & -0.246 &  &  &  &  &  &  & \tabularnewline
\hline 
\noalign{\vskip\doublerulesep}
\multirow{4}{*}{$\kappa_{\mathrm{sia}}$} &  & $w_{\kappa_{\mathrm{sia}},1}$ & $w_{\kappa_{\mathrm{sia}},2}$ & $w_{\kappa_{\mathrm{sia}},3}$ &  & $\sigma_{\kappa_{\mathrm{sia}}}$ &  &  & \tabularnewline
\cline{3-5} \cline{4-5} \cline{5-5} \cline{7-7} 
\noalign{\vskip\doublerulesep}
 &  & -0.008 & -0.005 & -0.025 &  & 0.490 &  &  & \tabularnewline
\noalign{\vskip\doublerulesep}
 &  & $b_{\kappa_{\mathrm{sia}}}$ &  &  &  &  &  &  & \tabularnewline
\cline{3-3} 
\noalign{\vskip\doublerulesep}
 &  & -0.494 &  &  &  &  &  &  & \tabularnewline
\hline 
\noalign{\vskip\doublerulesep}
\multirow{4}{*}{$r_{\mathrm{vac}}$} &  & $w_{r_{\mathrm{vac}}}$ &  &  &  & $\sigma_{r_{\mathrm{vac}}}$ &  &  & \tabularnewline
\cline{3-3} \cline{6-7} \cline{7-7} 
\noalign{\vskip\doublerulesep}
 &  & 12.944 &  &  &  & 0.279 &  &  & \tabularnewline
\noalign{\vskip\doublerulesep}
 &  & $b_{r_{\mathrm{vac}}}$ &  &  &  &  &  &  & \tabularnewline
\cline{3-3} 
\noalign{\vskip\doublerulesep}
 &  & -5.588 &  &  &  &  &  &  & \tabularnewline
\hline 
\noalign{\vskip\doublerulesep}
\multirow{4}{*}{$\kappa_{\mathrm{vac}}$} &  & $w_{\kappa_{\mathrm{vac}}}$ &  &  &  & $\sigma_{\kappa_{\mathrm{vac}}}$ &  &  & \tabularnewline
\cline{3-3} \cline{7-7} 
\noalign{\vskip\doublerulesep}
 &  & -1.299 &  &  &  & 0.516 &  &  & \tabularnewline
\noalign{\vskip\doublerulesep}
 &  & $b_{\kappa_{\mathrm{vac}}}$ &  &  &  &  &  &  & \tabularnewline
\cline{3-3} 
\noalign{\vskip\doublerulesep}
 &  & 1.446 &  &  &  &  &  &  & \tabularnewline
\hline 
\noalign{\vskip\doublerulesep}
\end{tabular}
}
\end{center}
\caption{\label{tab:h_model_num} Parameters defining the generative model of point defects generated by PKAs in hcp Zr. It is assumed that the quantities are expressed in $\mathrm{keV}$ and $\mathrm{\mathring{A}}$. The associated expectations, probabilities and transformation definitions are given in the table \ref{tab:h_model}. The parameters are provided in a form that can be used directly in the evaluation of the linear model of expectation $\mu$ (table \ref{tab:h_model}). Recall that in the domain where $\mu$ is a linear model, samples can be generated using Gaussian distributions and transformed "back" according to the mapping $f$.}
\end{table}

Here we would like to emphasise that one of the advantages of our simplified approach is that a single table makes for a fairly sophisticated descriptor of a cascade. Such compactness of representation would be rather difficult to achieve using non-parametric methods. For example, Gaussian process regression (GPR) or a deep neural network would result in matrices that are large and not easily transferable. An interesting side note is that in our attempts to use GPR we lost most of the correlations between features that we choose to assess.

\begin{figure}[h!]
\centering
\includegraphics[width=1.00\textwidth]{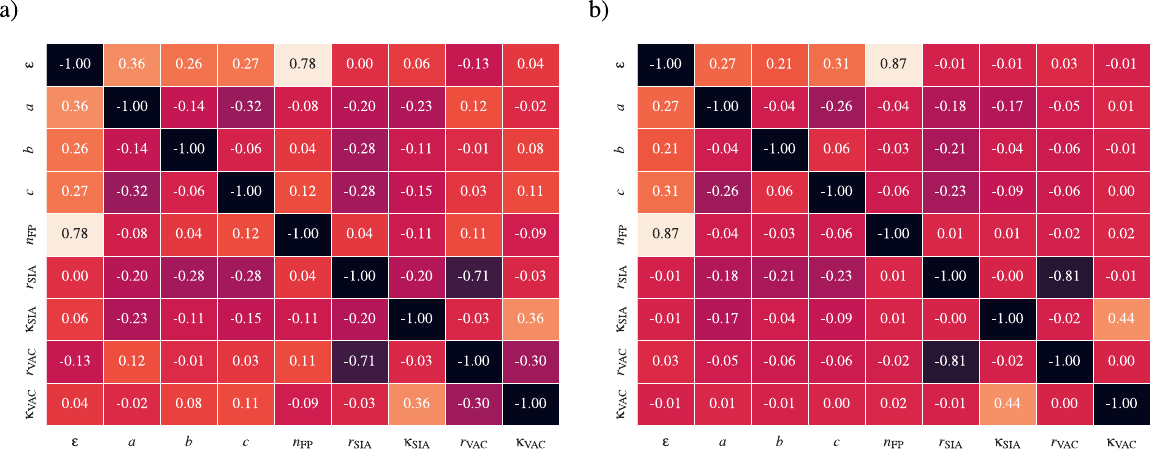}
\caption{\label{fig:org93fad12} Comparison between partial correlations of features of cascades from explicit MD simulations (a) and populations of defects obtained using the generative model (b). The quantities of interest are labelled in the same way as in figure \ref{fig:corr_heat_map} and are as follows: \( \epsilon \) -- the PKA energy, \( a,b,c \) the dimensions of the cascade, \(n_{\mathrm{FP}}\) -- the number of Frenkel pairs, \( r_{\mathrm{SIA/VAC}} \) -- average distance from the cluster centre for self interstitial atoms and vacancies, \( \kappa_{\mathrm{SIA/VAC}} \) -- concentration parameter for the von Mises-Fisher distribution.}
\end{figure}

We begin our assessment of the model with comparison of partial correlations. We compare cascades, or rather populations of defects, from explicit atomistic MD simulations and ones that were generated using the generative model. Results are presented in Figure \ref{fig:org93fad12}.
We were able to accurately recreate most of the correlations. However, because we impose dependencies before fitting, any unspecified correlations are lost, resulting in an overestimation of the remaining ones.

One might ask why we did not estimate a whole covariance matrix, of all parameters, from the data. The first reason is that the aim was to construct a relatively uncomplicated parametric model. Furthermore, when we tested a non-parametric method, namely the Gaussian Process Regression, even fewer partial correlations were recreated. For this reason, we decided that the optimal way forward was to develop the hierarchical model.

Having features of the cascade and the hierarchy of distributions, we can generate actual positions of defects. Strictly speaking, we will be sampling their approximate positions, as we use a continuous approximation. To obtain actual coordinates we need to shift them to the closest lattice point. However, for simplicty we ignore this requirement for now.

Before evaluating the performance of the model, we need to consider some final details. So far we have considered the average distance from the cluster centre, although we haven't said anything about the actual distribution. Analysis of the distances using kernel density estimation revealed that in many cases the distribution resembles a Gaussian distribution. This means that, in addition to the location parameter (e.g. the average), we also need a scale parameter. Because a Gaussian distribution is not appropriate for modelling distances that are strictly non-negative, and we also have a fairly large variance compared to the expectation, we decided that the best selection will be the truncated-normal distribution. Furthermore, we found close to no correlation between location and scale parameters. Note that we use a standard parametrisation where location \(\mu\) and scale \(\sigma\) correspond to expectation and standard deviation of the parent normal distribution. For that reason, we assume a constant parameter \(\sigma\), estimated from the data set, to be 0.619 for SIA and 0.519 for vacancies. Recall that, because we are working with PCA coordinates the actual spread will vary with the PKA energy.

In summary, we have defined a way to sample expected distances (local coordinates) of defects from the cluster centre as well as concentration parameters for associated von Mises distributions (separately defined for VAC and SIA). The hierarchical model also provides us with principle axes of the enclosing ellipsoid that define the shape of the cascade. These axes are used to transform samples from local coordinates to the space of a crystal. Based on the previous argument, the expected direction of concentration and the orientation of the enclosing ellipsoid are sampled uniformly over all orientations.

\begin{figure}[h!]
\centering
\includegraphics[width=1.00\textwidth]{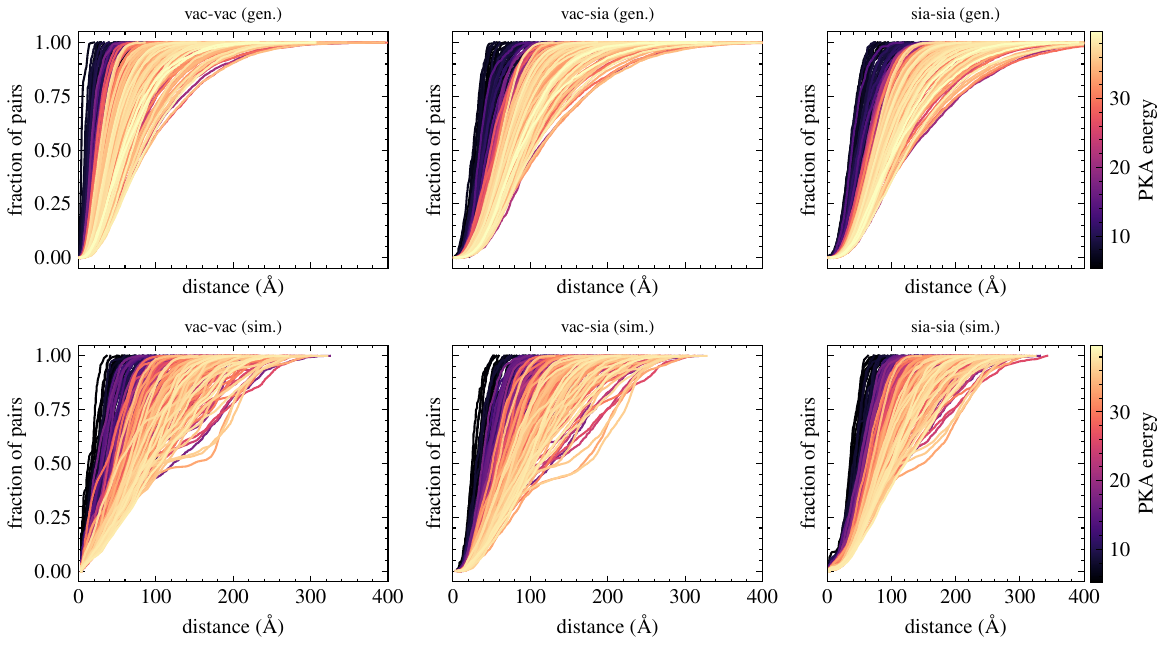}
\caption{\label{fig:def_pair_dist} Fraction of Frenkel pairs as a function of the search radius originating from the centre of the cluster of defects. The first row represents generated results, while the second -- MD simulations. Columns correspond to the type of a pair. \cmt{capitalise defects labels. Not sure about radial dist. yet. Good idea, but populations are small.}}
\end{figure}

To test the performance of our model we introduce a method inspired by the topological data analysis (TDA). We will count the number of connected pairs with an increasing "search" radius. However, instead of the actual number of connected defects, we will use fractions, Hence, we will get an analogue of a cumulative distribution function. This concept is somewhat related to barcodes in TDA. We have focused on distances between pairs as the key impact of the defect distribution in its role as the starting point for microstructural evolution via influencing chances of annihilating defects or forming clusters of vacancies or interstitials. This can be an important factor as diffusion happens on much longer time-scales than cascade evolution. The results are presented in Figure \ref{fig:def_pair_dist}.
We can see that the defects generated using the model correspond reasonably well to data generated using MD. The main difference is that some MD simulations have clearly much more complex structure. High-fidelity MD data contains some bimodal histograms of distances between pairs. This suggests some level of branching that our model is unable to recreate due to our consideration of a fixed number of features.

To make a more objective and numerical assessment, to each estimated curve we fit a cumulative distribution function of the Weibull distribution
\begin{equation}
F(x) = 1 - e^{-\left(\frac{x}{\lambda}\right)^k}.
\end{equation}
This gives us a simplified representation of a curve that consists of two parameters: characteristic scale \(\lambda\) and exponent \(k\). The distribution will enforce some properties (\textit{e.g.} unimodality), however, it  will still give us a fair representation of the considered relationship. 
\begin{figure}[h!]
\centering
\includegraphics[width=1.00\textwidth]{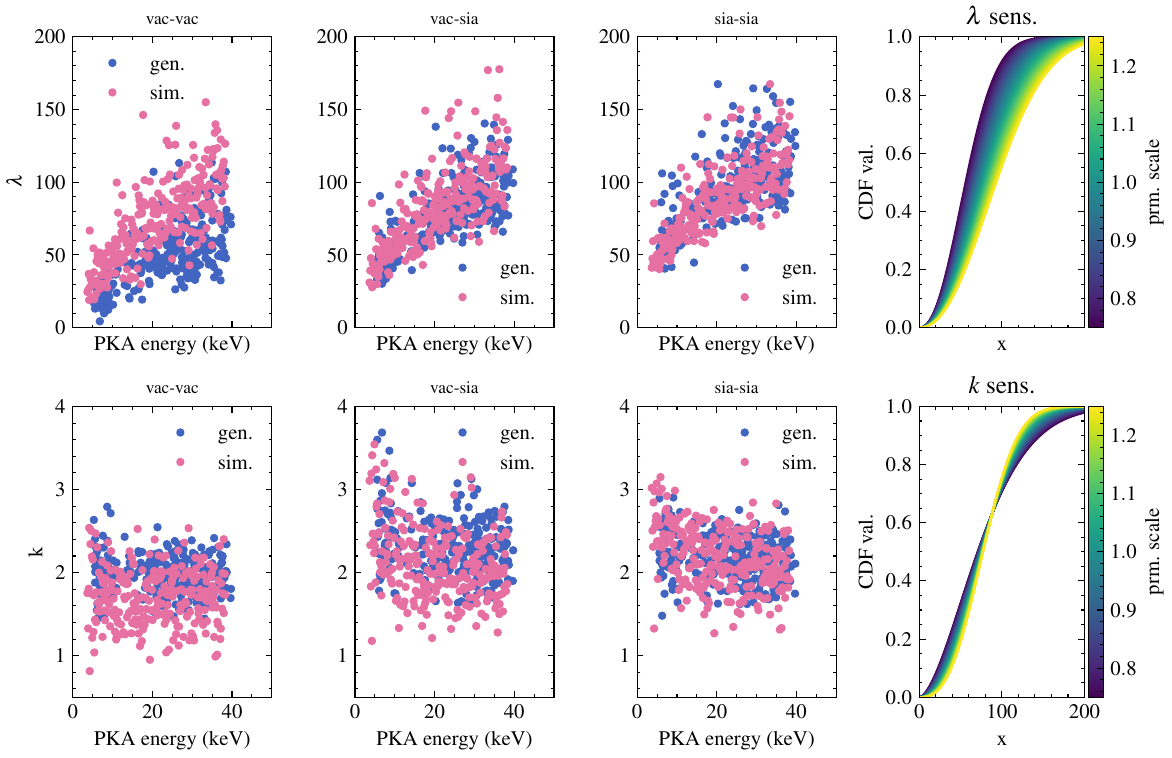}
\caption{\label{fig:org40dd8f9} Parameters of the Weibull distribution fitted to data from figure \ref{fig:def_pair_dist} (gen. -- data generated by the model, sim. -- results from MD simulations). The last column demonstrates sensitivity of the cumulative distribution function to the scale parameter $\lambda$ and the exponent $k$ (shape parameter). Each line of a given colour represents a scaled version $\lambda$ or $k$ with the base value being the average for the whole set.\cmt{I think}.}
\end{figure}
The results presented in Figure \ref{fig:org40dd8f9} demonstrate that almost all aspects of the distances between pairs are consistent between the model and the MD simulations. The key parameter here is \(\lambda\), which defines the scale of the distribution. We have good agreement between estimates for pairs of SIAs and for vacancies and SIAs. However, the distances between vacancies seem to be overestimated. On the other hand, due to the volatile nature of defect production, the final results may be difficult to distinguish. Which we find satisfactory.

We believe that the main issue is that we selected the truncated normal to represent the distance. Most likely a distribution that has zero probability density at zero distance (\textit{e.g.} log-normal) would be better in this particular case. However, at the moment we found the truncated-normal distribution the best compromise between reliability, complexity of implementation and representativeness.

\section{Baseline approximations}
\label{sec:base_approx}

In this section we will demonstrate the utility of the generative model in a case study. We have mentioned before that complete molecular dynamic simulations can be extremely expensive if we wish to reach out to experimental scales. However, we can use the generative model to construct a low-fidelity approximation that can provide some insights into the nature of radiation damage without making extremely extensive, and expensive simulations.

In the considered case study, we will demonstrate the use of the generative model to construct a baseline approximation - one that is sufficiently accurate to investigate the scale of an effect (\textit{e.g.} influence of the spectra on damage accumulation), assist in planning large-scale and multi-physics simulations (\textit{e.g.} estimating the number of simulations necessary to find the damage saturation point) or provide a starting point for higher-scale methods (\textit{e.g.} by populating a crystal with defects that are representative of a given dose for a given spectrum).

Our example can be considered as complementary to other low-fidelity approximations such as the binary collision approximation (BCA) or empirical relationships like NRT, ARC or RPA damage predictions \cite{nordlund_improving_2018}. Compared to these, our approach includes additional information about the spatial distribution of defects and the influence of temperature.  



Consider the common practice of reporting results of experiments and simulations using some variation of the dpa unit (displacement per atom). This sometimes leads to confusion and interpretation of this measure of dose as a measure of damage. However, this is necessary to make experiments and simulations comparable across different materials and irradiation types. Some researchers choose to use simple scaling using the NRT model. More often the measure of dose is obtained via BCA simulations, usually implemented in SRIM \cite{ZIEGLER20101818}. Nevertheless, the information is reduced to dose, i.e. energy transferred to a specific material.

It has been demonstrated by explicit MD simulations of build-up radiation (\textit{e.g.} \cite{tian_radiation_2021}) that dose and the type of material might be insufficient information to recreate the conditions of an experiment or a simulation. The complete information should also consist of some information about the PKAs spectrum.

Here, we would like to demonstrate that our generative model can be used to predict, at least qualitatively, the effect of the build-up without the extreme cost of running subsequent MD simulations. We choose to conduct three independent simulations with different PKA energies -- \(10 \mathrm{keV}\), \(20 \mathrm{keV}\) and \(30 \mathrm{keV}\). In each simulation, we introduce a ``cascade'' to a finite volume. Here, a ``cascade'' is a collection of point defects, vacancies and self-interstitial atoms, defined by their coordinates in the Cartesian space and generated using our model. This collection simulates the result of said collision ``cascade''. The centre of the cluster is sampled uniformly. Whenever a cascade is generated, we sample from its features, that then are used to simulate the positions of the point defects. Here, we use the same hierarchy, parameters and distributions that we selected and estimated in previous sections. The key assumption is that whenever a cascade is introduced, all preexisting defects are removed within the volume of the new cascade. This volume of space is defined by the principal axes. The mechanism is based on the consideration that collision cascades involve a thermal spike that will initiate a recovery. On the other hand, we are aware that this is just a crude approximation of a complex mechanism.

\begin{figure}[h!]
\centering
\includegraphics[width=1.00\textwidth]{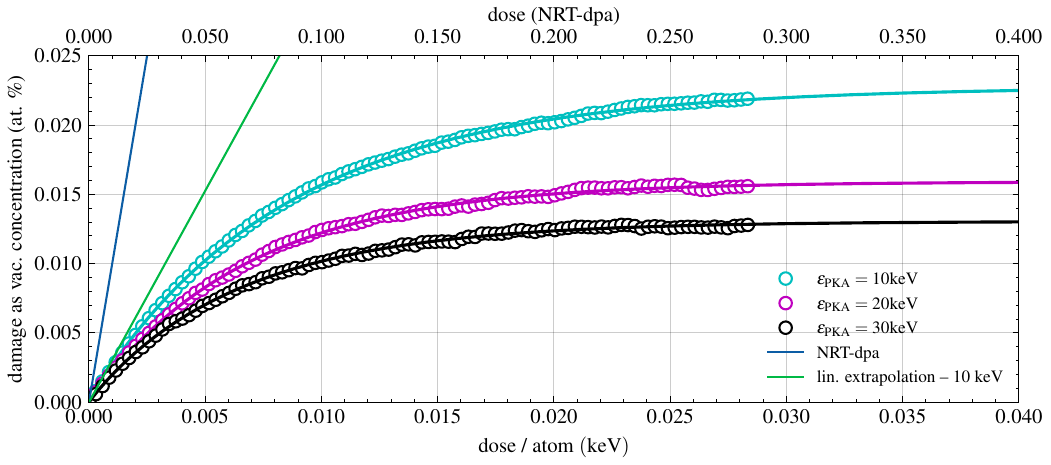}
\caption{\label{fig:org6bee910} Vacancy concentration (population size/number of atoms in the volume) as a function of dose. Predictions are made using an approximation based on the generative model. To provide a visual guide, the results are interpolated using the function: $a (1 - \exp(bx^c))$, where $a$, $b$ and $c$ are adjustable parameters. The threshold displacement energy for the NRT model was set to 40 eV.}
\end{figure}

The results are presented in Figure \ref{fig:org6bee910}.
It immediately transpires that we can observe a saturation of defects that is a direct consequence of introducing the annealing effect. While this is something we expected, it is not the main point of this simple experiment. In the limit of high dose delivered on a time-scale that allows us to ignore diffusion, we can observe that there is a significant difference between damage induced by PKAs with different energies. Even though in each case we used the same dose (defined as the PKA energy per atom in the simulation cell). A similar conclusion can be derived directly from the evaluation of the model parameters, which suggest that higher energy cascades result in lower defect density (no. of defects per unit volume, Figure \ref{fig:desc_res}). The conclusion is that the spectra of radiation will influence the nature of damage in a meaningful way. Therefore, the information about the dose alone is insufficient to infer the damage accurately, i.e. dose does not have a simple relationship to damage.

The above results are consistent with explicit MD simulation conducted by Tian \textit{et al.} in \cite{tian_radiation_2021}. However, quantitative comparison reveals the limit of this approach. In the simple model of damage saturation happens at a much slower rate (although the order of magnitude is correct) and accumulation of damage is overestimated at least by a factor of 2. We could adjust the rate if we chose a smaller interaction volume proportional to the size of a cascade, but this would increase the predicted limit of the damage. We could also remove only a fraction of defects under the new cascade and improve the agreement with explicit simulations. However, this is not the point we are trying to make.

The above case study shows that, even with very simple assumptions, the generative model can be used to draw important conclusions. We can build frameworks that can have an interpretable driving mechanism, such as the number of defects annealed due to the overlap of cascades, or it can provide an additional investigation tool.





For example, with some improvements to the presented framework, we can assess if changes in the spectra, like in the case of comparing samples from different reactors, should be taken into account. With the right parametrisation, it is just a matter of tweaking the first distribution in the model hierarchy. Furthermore, note that the BCA method does not involve the annealing effect and purely empirical models (like NRT) do not provide any spatial information.

Furthermore, the generative models, most likely including the simple model of damage accumulation, can also be used to introduce initial damage in simulations that investigate diffusion and/or formation of higher-order defects (e.g. dislocation loops). This would be more realistic than a random sample from a uniform distribution as multiple cascades will generate local concentration points even without the diffusion.


Furthermore, the generative model may enable us to establish a link between the number of defects introduced and the actual energy delivered to the system. This would allow direct comparison of results from simulations that rely on explicit insertion of defects, such as CRA/RDA \cite{derlet_microscopic_2020,warwick_microstructural_2021,maxwell_atomistic_2020}, with experiments, assuming we know the PKA spectrum. In other words, by becoming a part of a larger multi-scale, multi-method framework, the generative model may become an essential step in connecting damage and the dose.

How this could work in practice? As mentioned before, from Figure \ref{fig:org6bee910} we see that depending on the PKA energy, the same number of defects (volume is the same in each case) corresponds to a different dose. We could use regression analysis to find the number of defects associated with a given dose and define this relationship to be dependent on the parameters of the PKA sanctum. Alternatively, we could evaluate this relationship analytically. For example, we can approximate dose, as the energy delivered to the system, by writing a posterior predictive distribution
\begin{equation}
p_{V}\left(E\mid N\right)=\int_{\vec{\theta}}p\left(E\mid\vec{\theta}\right)p\left(\vec{\theta}\mid N\right)d\theta,
\end{equation}
where \(N\) is the number of defects introduced to the system and \(E\) is the amount of energy delivered. The vector \(\vec{\theta}\) represents parameters of the generative model, as well as parameters of the PKA spectrum. In the case of our generative model, the prior distributions of model parameters were taken as the Dirac delta (no ambiguity). However, we can include information about the degree of certainty we have in these parameters. The term \(p(E \mid \vec{\theta})\) will be essentially the spectrum, while \(p(\vec{\theta} \mid N)\) is factoring the particular form of the spectrum depending on the number of defects. The latter can be transformed using Bayes' theorem to a term that will include some form of the generative model.

\section{Summary and conclusions}
\label{sec:orgecd1f47}


For clarity, this section will summarise previously considered conclusions. The following are considered, in our view, to be the main contributions:
\begin{itemize}
\item We provide a database of collision cascades simulated with MD and the previously discussed TTM. This database can be used as a direct input to other methods, such as kMC, thanks to the well-designed sampling of PKA momenta. The database is accompanied by pre- and post-processing Python code showing how to access all the data behind the results presented in this paper. This allows them to be easily reproduced or to further the analysis. 
\item We developed a generative model for collision cascades. This model can be used to substitute/interpolate the associated database, \textit{i.e.} it can be used to generate representative populations of Frenkel pairs.
\end{itemize}
These key contributions are supplemented by the following:
\begin{itemize}
\item The key feature of the generative model is its low complexity. Given that it proved to be sufficiently representative, we believe that we have found a close-to-optimal method and family of distributions for interpolating databases of collision cascades.
\item We were able to overcome the problem of excessive heat in simulations with TTM using a simple Python extension of the input script. This was done by solving a simplified heat transfer equation and replacing TTM with a "single point" Langevin thermostat in the later stages of a cascade evolution.
\item We offer a practical parametrisation of the two-temperature model (TTM) for Zr, which can be efficiently incorporated into LAMMPS input files. Upon a reasonable request, we can provide the code we used to evaluate all quantities.

\item We have demonstrated how the generative model can facilitate a deeper understanding of radiation damage by providing a fast way to generate defects. This can be used to build baseline approximations that allow us to qualitatively evaluate expensive to explicitly simulate phenomena, such as radiation damage build-up and its relationship to the PKA spectrum, or the role of annealing due to cascade overlap.
\item As a result, we also developed a language of description, \textit{i.e.} a set of numerically representable and physically relevant features that can be easily published and shared. The usefulness of this descriptor is confirmed by a fairly successful comparison with actual simulated results of collision cascades.
\item We have developed a technique to represent and compare clouds of point defects. We use cumulative distributions that represent the fraction of defects within a specific radius. This representation has a direct relationship with minimum diffusion paths between pairs of defects. This approach can be applied to analyse other characteristics in a manner similar to techniques used in the topological data analysis.
\end{itemize}

The methodology we have presented has a potential for further development. For example, we can use the Bayesian framework to improve the comparability of different studies in the domain of radiation damage and to build a network of conditional probabilities. Note that the generative model essentially represents the relationship between the quanta of the radiation spectrum and the defects introduced by collision cascades. We are therefore opening up the possibility of relating the damage to the dose, using the information about the spectra. The Bayesian framework may allow us to propagate the information in both directions in a multi-scale, multi-method framework.

There are, nevertheless, a number of ways in which this study could be improved. For example, we might extend the descriptor of a cascade by including some information about the spread of defects within the local coordinates (currently it is a fixed value). We could also take advantage of non-parametric modelling and discover more nuanced relationships between features. We believe that heteroscedastic Gaussian process regression with imposed hierarchy could be an alternative to the proposed model. However, we would lose the advantages of an explicit model. 

The model we have developed is also not complete. By design it is based on collision cascades up to \(40 \, \mathrm{keV}\). Hence we use only half of the PKA spectrum for Zr ($n^0$ flux) in representative reactor conditions. It is expected that higher energy cascades will branch out. While the replacement of high energy cascades by multiple lower energy cascades may be acceptable, it should be considered as a very simple approximation. The real picture is more complicated. Although it has been shown that they do indeed tend to branch out, these events may involve interconnected cascades \cite{zhou_molecular_2018}. Even in the range of PKA energies we considered, we observed bimodal distributions of pairwise distances, suggesting the presence of branch-out cascades. Therefore, future improvements should take into account the emergence of such structures.

The last point we want to address is that we did not consider the directions of PKAs because we focused on high temperature simulations (\(600 \, \mathrm{K}\)). Our analysis suggests that this is not an issue at these temperatures. However, at temperatures close to \(0 \, \mathrm{K}\) it could be an important factor.

\section{Data and code availability}

The database of pre- and post-processing scripts, input files and results from molecular dynamics simulations used in this study will be publicly available on the Zenodo platform \cite{barzdajn_2024_10554641}. This includes examples of Python code demonstrating how to access and interpret the data. Furthermore, a code example that implements the generative model from this paper  will also be published in an open-access format. 

\section{Acknowledgements}

We would like to kindly acknowledge The Engineering and Physical Sciences Research Council (EPSRC) for funding the MIDAS project (Mechanistic understanding of Irradiation Damage in fuel Assemblies -- ref. EP/S01702X/1). C P Race was funded by a University Research Fellowship of the Royal Society. Calculations were performed on a computational cluster, maintained by the Computational Shared Facility, The University of Manchester.

\section{Author contributions}

\textbf{Bartosz Barzdajn}: Writing - Original draft, Conceptualization, Methodology, Software, Validation, Formal analysis, Investigation, Data curation, Visualization. \textbf{Christopher Race}: Writing - Review \& Editing, Conceptualization, Methodology, Validation, Formal analysis, Supervision, Project administration, Funding acquisition.

\Urlmuskip=0mu plus 1mu
\section{References}
\label{sec:org1ef5b6f}

\bibliographystyle{unsrturl}
\bibliography{software,collision_cascades,data}
\end{document}